\documentclass[aps,floats]{revtex4}
\usepackage{amsmath,amssymb}
\usepackage{graphicx,epsfig}

\begin{document}
\bibliographystyle {plain}

\def\oppropto{\mathop{\propto}} 
\def\opsimeq{\mathop{\simeq}}
\def\opoverderline{\mathop{\overline}}
\def\operarrow{\mathop{\longrightarrow}}
\def\opsim{\mathop{\sim}}

\def\fig#1#2{\includegraphics[height=#1]{#2}}
\def\figx#1#2{\includegraphics[width=#1]{#2}}


\title{ On the critical weight statistics of the Random Energy Model \\
and of the Directed Polymer on the Cayley Tree } 


\author{ C\'ecile Monthus and Thomas Garel }
 \affiliation{Service de Physique Th\'{e}orique, CEA/DSM/SPhT\\
Unit\'e de recherche associ\'ee au CNRS\\
91191 Gif-sur-Yvette cedex, France}

\begin{abstract}
We consider the critical point of two mean-field disordered models : (i) 
 the Random Energy Model (REM), introduced by Derrida as a mean-field
spin-glass model of $N$ spins
(ii)  the Directed Polymer of length $N$ on a Cayley Tree (DPCT)
with random bond energies.
Both models are known to exhibit a freezing transition
between a high temperature phase where the entropy is extensive 
and a low-temperature phase of finite entropy, where the weight statistics 
coincides with the weight statistics of L\'evy sums with index $\mu=T/T_c<1$.
In this paper, we study the weight statistics at criticality via the
entropy $S=-\sum w_i \ln w_i$
and the generalized moments $Y_k=\sum w_i^k$, 
where the $w_i$ are the Boltzmann weights of the $2^N$ configurations.
In the REM, we find that the critical weight statistics is governed 
by the finite-size exponent $\nu=2$ : the entropy scales as $\overline{S}_N(T_c) \sim N^{1/2}$,
the typical values $e^{\overline{ \ln Y_k}}$ decay as $N^{-k/2}$, and 
the disorder-averaged values $\overline{Y_k}$
are governed by rare events and decay as $N^{-1/2}$ for any $k>1$.
For the DPCT, we find that the entropy scales similarly as $\overline{S}_N(T_c) \sim N^{1/2}$,
whereas another exponent $\nu'=1$ governs the $Y_k$ statistics :
the typical values $e^{\overline{ \ln Y_k}}$ decay  as $N^{-k}$,
the disorder-averaged values $\overline{Y_k}$ decay as $N^{-1}$ for
any $k>1$. As a consequence, the asymptotic probability distribution
$\overline{\pi}_{N=\infty}(q)$ of the overlap $q$,
beside the delta function $\delta(q)$ which bears the whole normalization,  
contains an isolated point at $q=1$, as a memory of the delta peak $(1-T/T_c) \delta(q-1)$ of the
low-temperature phase $T<T_c$. The associated value
$\overline{\pi}_{N=\infty}(q=1)$ is finite for the DPCT, and diverges as
$\overline{\pi}_{N=\infty}(q=1) \sim N^{1/2}$ for the REM.

\bigskip

PACS numbers: 75.10.Nr, 02.50.-r, 05.40.Fb

\end{abstract}

\maketitle

\section{ Introduction}

Spin-glasses \cite{Bin_You,replica} and directed polymers in random media
 \cite{Hal_Zha} 
are two kinds of disordered models where the relations between 
 finite dimensional models and mean-field models have remained controversial
 over the years.
For the directed polymer in a random medium of dimension $1+d$, with $d=3$
where a localization/delocalization transition occurs, we have recently found
numerically that the weight statistics is multifractal
at criticality \cite{DP3dmultif},  in close analogy with models
 of the quantum Anderson localization transition, where 
the multifractality of critical wavefunctions is well established
\cite{andersonloc,Mirlin}. 
In this paper, our aim is to study the critical weight statistics 
 in the mean-field version of the model,
namely the Directed Polymer on a Cayley Tree (DPCT) \cite{Der_Spo,Coo_Der}.
This model presents many similarities \cite{Der_Spo,Coo_Der} with
the Random Energy Model (REM), introduced by Derrida as a mean-field
spin-glass model \cite{rem} :
both models are known to exhibit a freezing transition
between a high temperature phase where the thermodynamic observables
coincide with their extensive annealed values,
and a low-temperature phase of finite entropy, in which the weight
statistics \cite{Der_Tou,Der_Fly} coincides with the weight statistics
of L\'evy sums with index $\mu=T/T_c<1$ \cite{Der}.
Therefore we also study the weight statistics of the REM at criticality,
as well as in L\'evy sums at the critical value $\mu_c=1$,
to compare with the results for the Directed Polymer on a Cayley Tree.
We find that the three models have different critical finite-size properties.

The paper is organized as follows.
In Section \ref{models}, we recall the main properties of the Random Energy Model (REM)
and of the Directed Polymer on a Cayley Tree (DPCT).
We then study in parallel the weight statistics at criticality for the two models :
we describe the properties of the entropy in Section \ref{entropie}, 
the decay of disorder-averaged values $\overline{Y_k}$ in Section \ref{ykav},
and the decay of  typical values $Y_k^{typ} = e^{\overline{ \ln Y_k}}$ in Section \ref{yktyp}.
To explain the differences between averaged and typical values,
we discuss in Section \ref{histo} the probability distributions
of the maximal weight $w_{max}$ and of $Y_2$ over the samples,
with a special emphasis on the rare events that govern averages values.
The Section \ref{overlap} is devoted to the finite-size properties
of the overlap distribution at criticality.
Section \ref{conclusion} contains our conclusions.
For clarity, the weight statistics of L\'evy sums is discussed
separately in the Appendices :
we first recall in Appendix \ref{app_mu} the results for $\mu<1$,
and we describe the critical case $\mu_c=1$ in Appendix \ref{app_muc}.

\section{ Models and observables }

\label{models}

\subsection{Reminder on the Random Energy Model }

The Random Energy Model (REM), introduced in the context of spin glasses
\cite{rem}, is defined by the partition function 
\begin{equation}
Z_N = \sum_{i=1}^{2^N} e^{- \beta E_{i}}
\label{zrem}
\end{equation}
where the energies $E_{i}$ of the 
$2^N$ configurations of $N$ spins 
are assumed to be independent random variables
drawn from the Gaussian distribution
\begin{equation}
P_N(E) = \frac{1}{\sqrt{\pi N}} e^{ -\frac{E^2}{N}}
\label{PErem}
\end{equation}
This model presents a freezing transition at \cite{rem}
\begin{eqnarray}
T_c= \frac{1}{2  \sqrt{ \ln 2}}
\label{remtc} 
\end{eqnarray}
and we now briefly recall the main properties
of the high temperature and low temperature phases.

\subsubsection{ High-temperature phase }

In the high temperature phase $T \geq T_c$, the free-energy
per spin coincides with the annealed free-energy \cite{rem}
\begin{eqnarray}
f(T>T_c)= f_{ann}(T) = - T \ln 2 - \frac{1}{4 T}
\label{remfannealed} 
\end{eqnarray}
As a consequence, the entropy per spin vanishes linearly
at the transition
\begin{eqnarray}
s(T>T_c)= s_{ann}(T) =  \ln 2 - \frac{1}{4 T^2}
 \oppropto_{T \to T_c^+ } (T-T_c)
\label{remsannealed} 
\end{eqnarray}
and the specific heat per spin remains finite
\begin{eqnarray}
c(T>T_c) = c_{ann}(T) = \frac{1}{2 T^2}
 \opsimeq_{T \to T_c^+ } c(T_c^+)=2 \ln 2
\label{remcannealed} 
\end{eqnarray}

\subsubsection{ Low-temperature phase }

From the thermodynamic point of view, the low-temperature
phase $T<T_c$ is completely frozen, with a constant free-energy per spin
\cite{rem}
\begin{eqnarray}
f(T \leq T_c) = - \sqrt{ \ln 2 }
\label{remflow} 
\end{eqnarray}
As a consequence, the entropy per spin 
vanishes in the whole low-temperature phase
\begin{eqnarray}
s(T \leq T_c)= 0
\label{remslow} 
\end{eqnarray}
as well as the specific heat per spin
\begin{eqnarray}
c(T<T_c)= 0
\label{remclow} 
\end{eqnarray}

To understand better the properties of this low-temperature phase,
it is convenient to study the statistical properties of the 
configurations weights in the partition function (Eq. \ref{zrem})
\begin{equation}
w_{i} = \frac{ e^{- \beta E_{i}} }{Z_N}
\label{defwi}
\end{equation}
It turns out that their statistics 
is in direct correspondence with L\'evy sums of index $0<\mu=\frac{T}{T_c}<1$  \cite{Der}
(see Appendix \ref{app_mu} for more details).
In particular, the moments
\begin{equation}
Y_k=\sum_{i=1}^{2^N} w_{i}^k
\label{defyk}
\end{equation}
have for disorder-averages \cite{Der}
\begin{equation}
\overline{Y_k}= \frac{\Gamma(k-\mu(T))}{\Gamma(k) \Gamma(1-\mu(T))}
\ \ {\rm  with } \ \ \mu(T)=\frac{T}{T_c}
\label{ykrem}
\end{equation}
The density $f(w)$ giving rise to these moments
\begin{equation}
\overline{Y_k} = \int_0^1 dw w^k f(w)
\label{lienykf}
\end{equation}
reads  \cite{Der}
\begin{equation}
f(w)= \frac{w^{-1-\mu} (1-w)^{\mu-1} }{ \Gamma(\mu) \Gamma(1-\mu)}
\label{densitew}
\end{equation}
and represents the averaged number of terms of weight $w$.
This density is non-integrable as $w \to 0$, because in the limit
 $N \to \infty$, the number of terms of vanishing weights diverges.
The normalization corresponds to 
\begin{equation}
\overline{Y_{k=1}} = \int_0^1 dw w f(w) =1
\end{equation}

In particular, as the transition is approached $\mu=T/T_c \to 1^-$
these disorder-averaged moments all vanish linearly for $k>1$
\begin{equation}
\overline{Y_k} \oppropto_{T \to T_c^- } (T_c-T)
\label{ykremtotc}
\end{equation}

The link between these weights properties and
 the thermodynamics is via the total entropy
 \cite{Gro_Mez}
\begin{equation}
S_N= - \sum_{i=1}^{2^N} w_{i} \ln (w_{i})
 = - \left [\partial_k \sum_{i=1}^{2^N} w_{i}^k \right]_{k \to 1}
= - \left [\partial_k Y_k \right]_{k \to 1}
\label{entropyGM}
\end{equation}
From Eq. \ref{ykrem}, the disorder-averaged value over the samples reads
\begin{equation}
\overline{S}_N(T<T_c)
= - \left [\partial_k \overline{Y}_k \right]_{k \to 1}
= \Gamma'(1)- \frac{\Gamma' \left( 1- \mu(T) \right)}
{\Gamma \left( 1- \mu(T) \right)}
\label{entropyGMres}
\end{equation}
In the critical region $ T \to T_c^-$, the entropy per spin 
presents the following finite-size scaling behavior
\begin{equation}
\overline{ s_N(T) } \equiv \frac{\overline{ S_N(T) } }{N} \oppropto_{T \to T_c^- } \frac{1}{N (T_c-T) }
\label{sremfsslow}
\end{equation}
Similarly, the disorder-averaged specific heat per spin follows 
\begin{equation}
\overline{ c_N(T) } \equiv \frac{\overline{C_N(T)}}{N}  \oppropto_{T \to T_c^- } \frac{1}{N (T_c-T)^2 }
\label{cremfsslow}
\end{equation}
These finite-size scaling behaviors are in agreement with
the more detailed finite-size corrections of the free-energy 
computed in \cite{rem}.

\subsection{Reminder on the directed polymer on a Cayley tree}

The directed polymer on a Cayley tree with disorder
has been introduced in \cite{Der_Spo}
 as a mean-field version
of the directed polymer in a random medium \cite{Hal_Zha}.
The model is defined by the partition function
\begin{eqnarray}
Z_N= \sum_{{\cal C}} e^{- \beta E({\cal C})}
\label{zcayley} 
\end{eqnarray}
where the $2^N$ configurations $\cal C$ are the paths of $N$ steps
on a Cayley tree with
coordination number $K=2$. The energy $E({\cal C})$ of a path
is the sum of the energies of the visited bonds.
Each bond has a random energy
drawn independently, for instance with the Gaussian distribution
\begin{eqnarray}
\rho(\epsilon) = \frac{1}{\sqrt{2 \pi}} e^{- \frac{\epsilon^2}{2} }
\label{rhocayley} 
\end{eqnarray}
As in Eq. \ref{defwi}, it is convenient to consider the 
configurations weights $w({\cal C}) = e^{- \beta E( {\cal C})}/Z_N  $
 in the partition function (Eq. \ref{zcayley})
and the associated moments $Y_k$ (Eq. \ref{defyk}).

\subsubsection{ Similarities with the Random Energy Model}

This model presents many similarities \cite{Der_Spo,Coo_Der} with the Random Energy Model
described above.
It presents a freezing transition at
\begin{eqnarray}
T_c= \frac{1}{ \sqrt{ 2 \ln 2 } }
\label{tccayley}
\end{eqnarray}
The free energy per step coincides with the annealed free energy above $T_c$
and is completely frozen below
\begin{eqnarray}
f(T) && =f_{ann}(T) = - T \ln 2 - \frac{1}{2 T} {\rm \ \ \ for \ \ }  T \geq
T_c \\
f(T) && = - \sqrt{ 2 \ln 2}   {\rm \ \ \ for \ \ }  T \leq T_c
\label{fcayley}
\end{eqnarray}
So at the thermodynamic level, all properties are the same
as in the REM : the entropy per step vanishes linearly
 at the transition (Eq. \ref{remsannealed}), and the specific heat per step
presents a jump (Eq. \ref{remcannealed}).
From the finite-size behavior of the free-energy for $T<T_c$
(Eq. 76 of \cite{Coo_Der}), one obtains by differentiation
with respect to the temperature that the entropy per step
and the specific heat per step have the same finite-size
scaling as in the REM (Eqs \ref{sremfsslow} and \ref{cremfsslow}  )
\begin{eqnarray}
\overline{ s_N(T) } && \equiv \frac{\overline{S_N(T)}}{N} \oppropto_{T \to T_c^- } \frac{1}{N (T_c-T) } \\
\overline{ c_N(T) } && \equiv \frac{\overline{C_N(T)}}{N}  \oppropto_{T \to T_c^- } \frac{1}{N (T_c-T)^2 }
\end{eqnarray}

It turns out that even beside these thermodynamic quantities,
the two models are still very similar.
In \cite{Der_Spo}, it was shown that, in the limit $N \to \infty$,
the distribution of the overlap $q$ between two walks
on the same disordered tree is simply the sum of
two delta peaks at $q=0$ and $q=1$ in
the whole low-temperature phase \cite{Der_Spo} : 
\begin{eqnarray}
\pi (q)= (1-Y_2) \delta(q) +Y_2 \delta(q-1)
\label{piqlow}
\end{eqnarray}
And the distribution of $Y_2$ over the samples is 
exactly the same as in the Random Energy Model \cite{Der_Spo}.
In particular, its averaged value reads \cite{Der_Spo}
\begin{eqnarray}
\overline{ Y_2(T) }= 1 - \frac{T}{T_c}
\label{y2av}
\end{eqnarray}

\subsubsection{ Differences with the Random Energy Model}

As explained in details in \cite{Coo_Der},
the differences with the Random Energy Model (REM)
is that the directed polymer on a Cayley tree
corresponds to a Generalized Random Energy Model (GREM)
with $p=N$ levels, whereas the REM corresponds
to the case of a GREM with $p=1$ level.
This induces some differences for the finite-size
properties of the free-energy \cite{Coo_Der}.
The conclusion of \cite{Coo_Der} is that 
the finite-size scaling behavior of the REM
only involves the product $(T_c-T) N^{1/{\nu}}$ with
\begin{eqnarray}
\nu=2
\label{nu}
\end{eqnarray}
whereas for the Cayley tree, the situation is more subtle,
and the product $(T_c-T) N^{1/{\nu'}}$ with another exponent
\begin{eqnarray}
\nu'=1
\label{nuprime}
\end{eqnarray}
also appears.
As a consequence, even if the weight statistics is the same
in the low-temperature phase of the two models, 
their critical properties might be different,
as we will indeed find in the following.

\subsection{ Numerical details}

The numerical results given in the following sections
for the REM and the DPCT have been obtained
for the following sizes $N$ (with $2^N$ configurations)
and the corresponding numbers $n_s(N)$ of disordered samples 
\begin{eqnarray}
N && = 5-14,16,18,20 \\
n_s(N) && = 10^7 , 4 \cdot 10^6 ,10^6 , 4 \cdot 10^5
\label{nume}
\end{eqnarray}

\section{ Entropy at criticality }

\label{entropie}

\subsection{ Disorder averaged entropy at criticality}

We have recalled in the previous Section
that both for the Random Energy Model
and for the directed polymer on the Cayley tree,
the freezing transition corresponds to a jump
in the intensive specific heat
\begin{eqnarray}
\overline{ c_N(T>T_c) } && \opsimeq_{T \to T_c^+} Constant \\
\overline{ c_N(T<T_c) } && \opsimeq_{T \to T_c^-} \frac{1}{N (T_c-T)^2}
\end{eqnarray}
The finite-size scaling thus involves the exponent $\nu=2$
\begin{eqnarray}
\overline{ c_N(T) } && = {\cal C} \left( (T_c-T)N^{1/\nu}\right)
 \ \ {\rm \ \ with \ \ }
\nu=2
\end{eqnarray}
For the REM, the explicit form of the scaling function can be obtained
from Eq. (64) of \cite{Coo_Der}.

Similarly, the total disorder-averaged entropy has the following
behaviors on both sides of the transition
\begin{eqnarray}
\overline{ S_N(T>T_c) } && \opsimeq_{T \to T_c^+} N (T-T_c) \\
\overline{ S_N(T<T_c) } && \opsimeq_{T \to T_c^-} \frac{1}{T_c-T}
\end{eqnarray}
One thus expect the following finite-size scaling form
\begin{eqnarray}
\overline{ S_N(T) } && = N^{1/2} {\cal S} \left( (T_c-T) N^{1/\nu}\right)
 \ \ {\rm \ \ with \ \ } \nu=2
\end{eqnarray}
At criticality, we thus expect both for the REM and for the Directed Polymer on the Cayley tree
\begin{eqnarray}
\overline{ S_N(T_c) } && \propto N^{1/2} 
\label{entropietc} 
\end{eqnarray}
in agreement with our numerical simulations for both models.

\subsection{ Entropy distribution at criticality}

\begin{figure}[htbp]
\includegraphics[height=6cm]{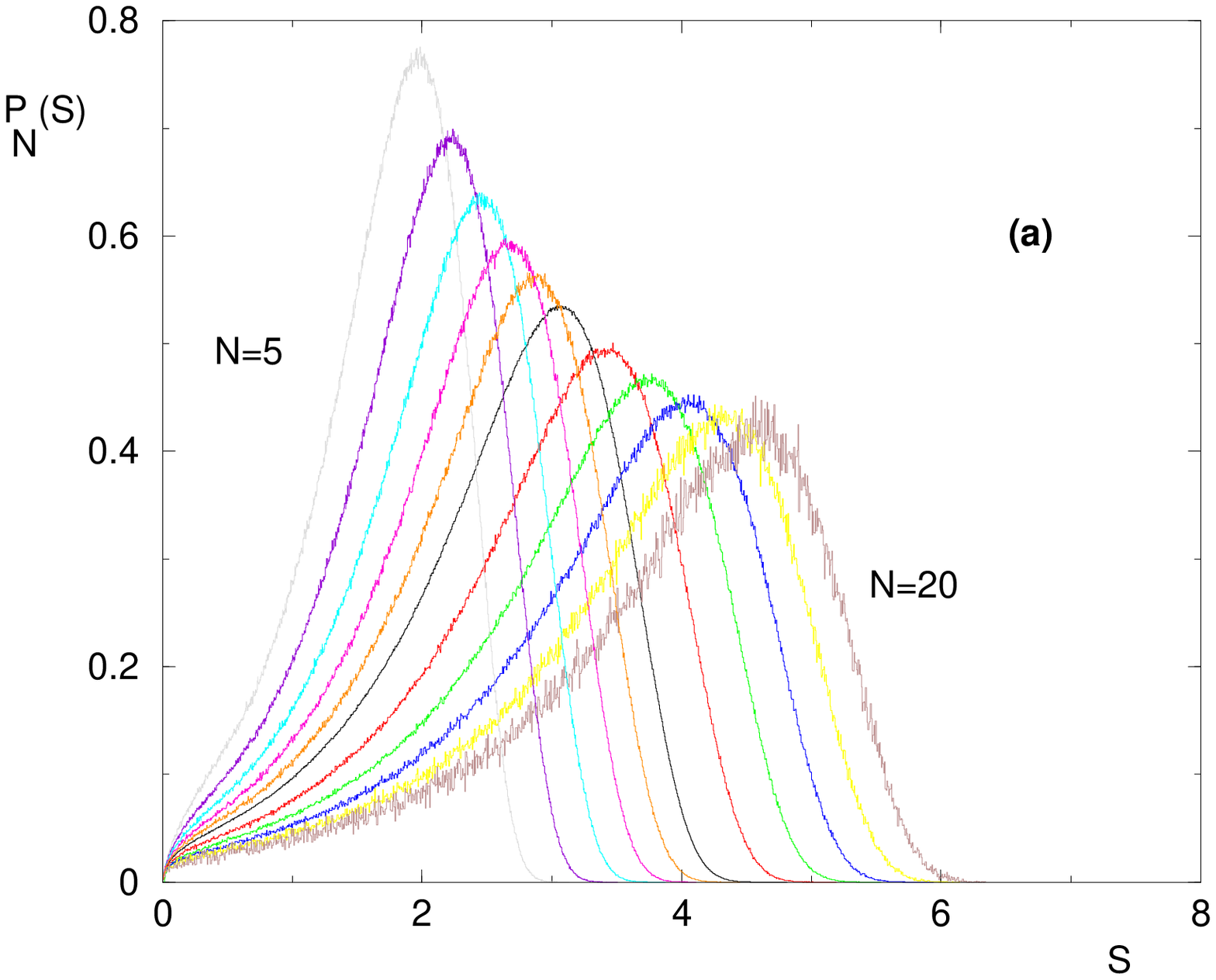}
\hspace{1cm}
\includegraphics[height=6cm]{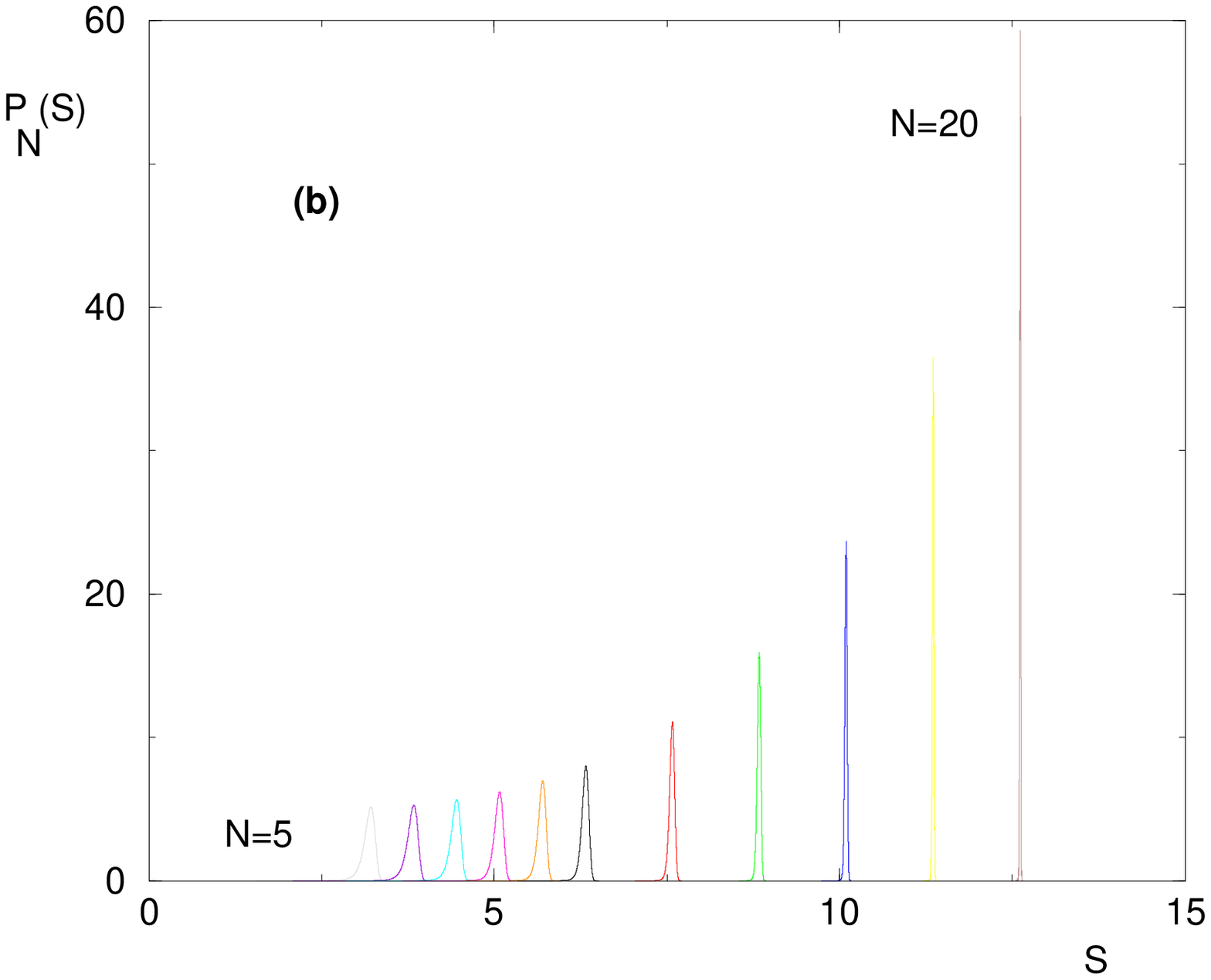}
\caption{ (Color on line) REM : entropy probability distribution $P_N(S)$
for sizes $N=5,6,7,8,9,10,12,14,16,18,20$
(a) at $T_c$, the distribution remains broad around the
averaged value $\overline{S} \propto N^{1/2}$, with 
a slow decay of rare events of small entropy $S \sim 0$.
 (b) at $T=2.>T_c$  : the width around the average value
$\overline{S}=N s_{ann}(T)$ decays exponentially in $N$,
in agreement with Ref. \cite{Bov} }
\label{sremhisto}
\end{figure}

\begin{figure}[htbp]
\includegraphics[height=6cm]{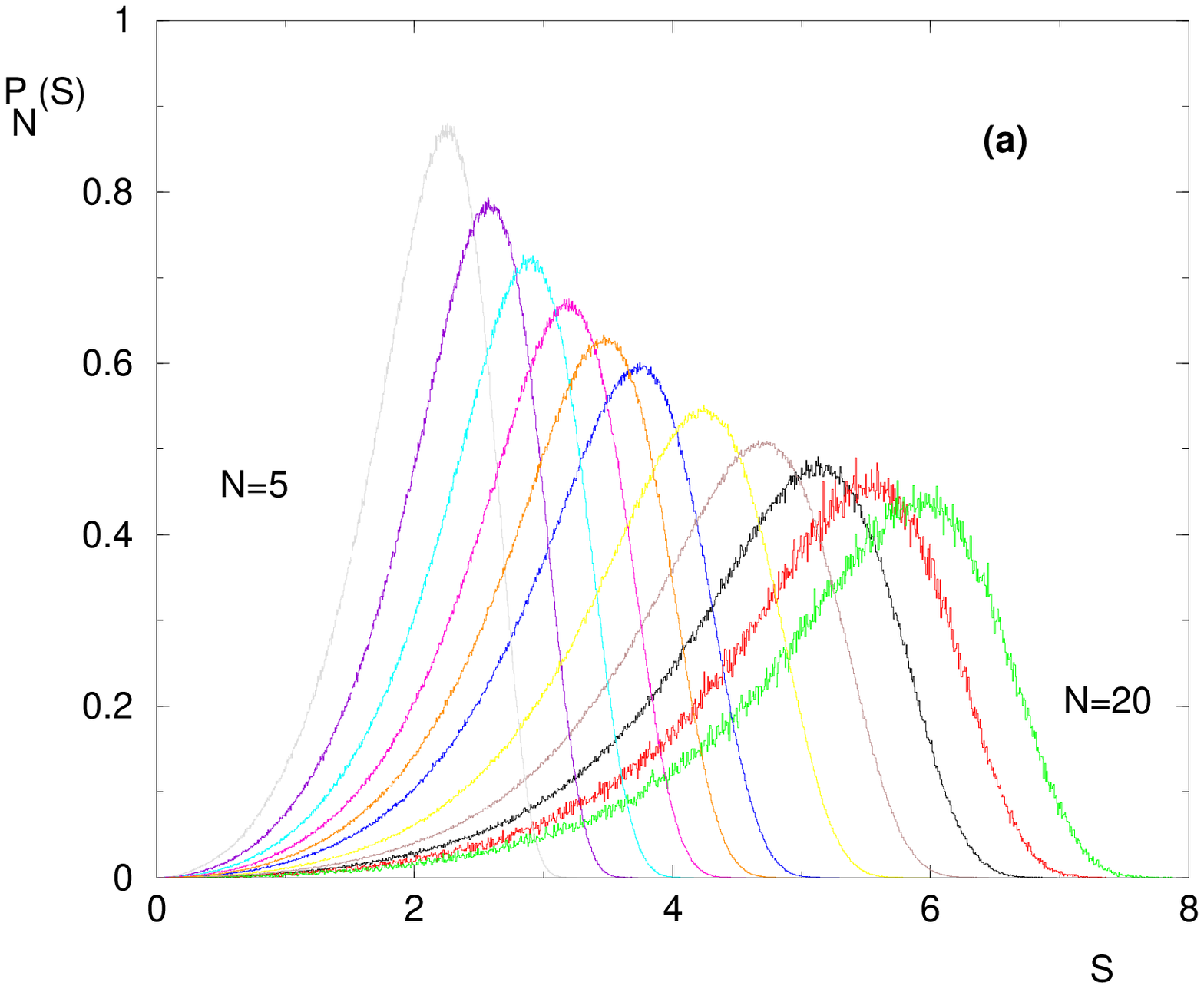}
\hspace{1cm}
\includegraphics[height=6cm]{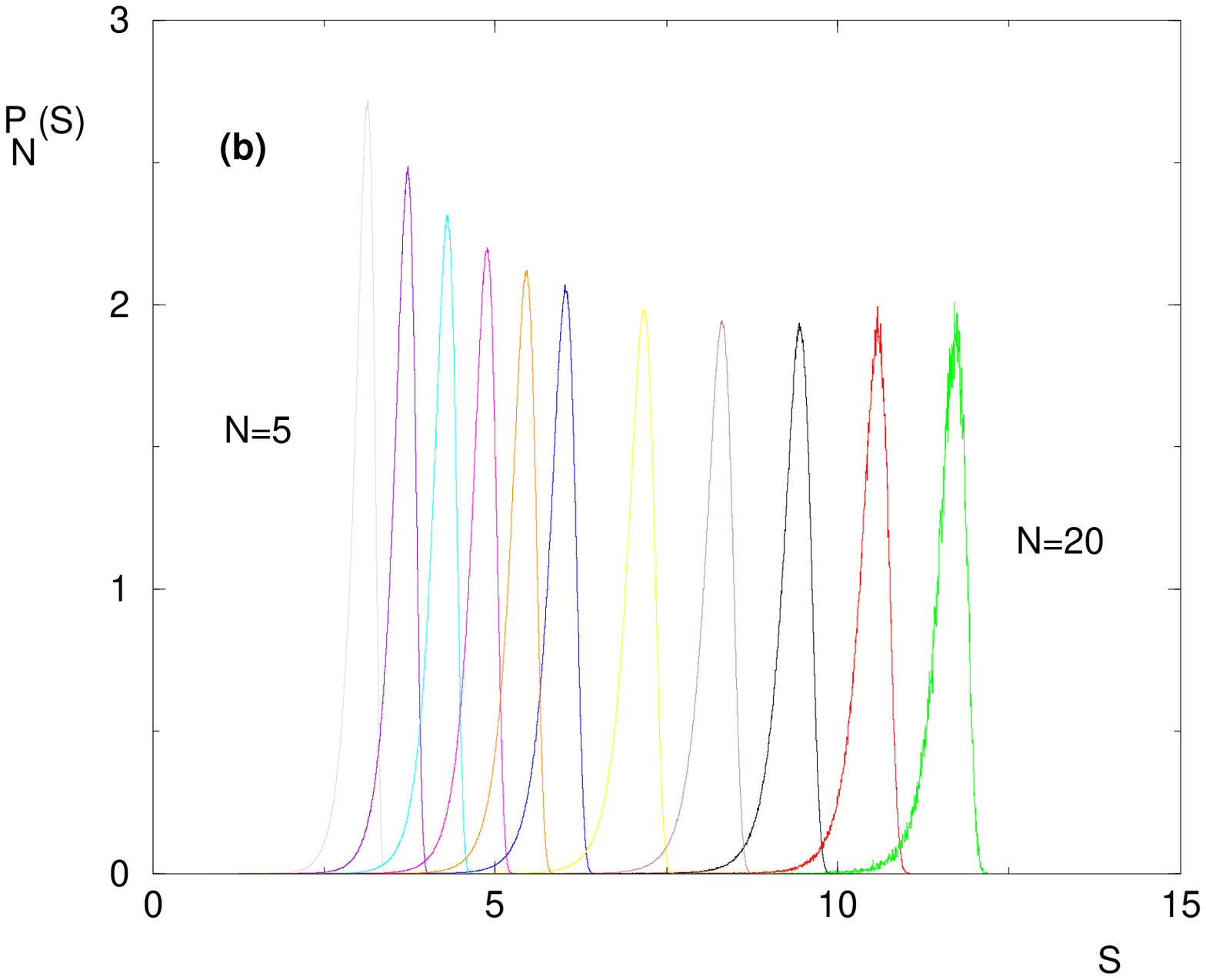}
\caption{ (Color on line) DPCT : entropy probability distribution $P_N(S)$
for sizes $N=5,6,7,8,9,10,12,14,16,18,20$
(a) at $T_c$, the distribution remains broad around the
averaged value $\overline{S} \propto N^{1/2}$. The
statistics of rare events in the region $S \sim 0$ is
different from the REM (see Fig. \ref{sremhisto} a ).
 (b) at $T=2.>T_c$ : the width around the average value
$\overline{S}=N s_{ann}(T)$ converges towards a constant,
in contrast with the REM (see Fig. \ref{sremhisto} b ). }
\label{scayleyhisto}
\end{figure}

We shown on Fig. \ref{sremhisto} and \ref{scayleyhisto}
 the probability distribution $P_N(S)$
of the entropy $S$ for the REM and for the DPCT respectively,
both at criticality and in the high-temperature phase for comparison.
At criticality, $P_N(S)$  remains broad around the
averaged value $\overline{S} \propto N^{1/2}$, with 
a slow decay of rare events of small entropy $S \sim 0$.
The comparison of Figs \ref{sremhisto} (a) and \ref{scayleyhisto} (a)
show that these rare finite samples that are still 'frozen'
at $T_c$ do not obey the same statistics in the REM
and in the DPCT (see the more detailed discussion
 on rare events in section \ref{rare}). 
In the high temperature phase, the width around the average value
$\overline{S}=N s_{ann}(T)$ decays exponentially in $N$ in the REM \cite{Bov},
as shown on Fig. \ref{sremhisto} (b), whereas it 
 converges towards a constant in the DPCT, as shown on Fig.
 \ref{scayleyhisto} (b).

\section{  Decay of disorder averaged values $\overline{Y_k}(N) $ 
at criticality }

\label{ykav}

\subsection{ Special case $k=2$ }

\begin{figure}[htbp]
\includegraphics[height=6cm]{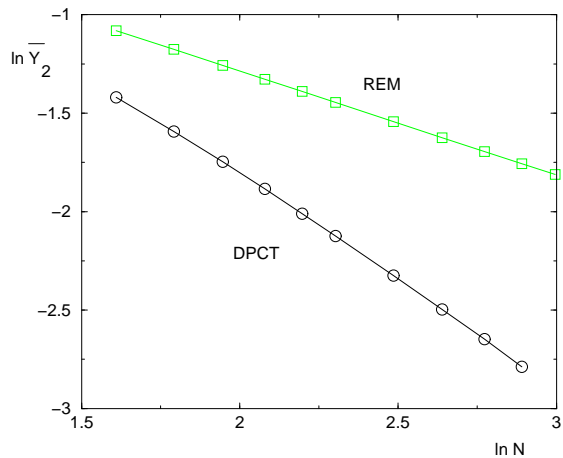}
\hspace{1cm}
\caption{ (Color on line) At criticality, the slope of $\ln \overline{Y_2}(N,T_c)$
 as a function of $(\ln N)$ (here $5 \leq N \leq 20$)
is of order $1/\nu =0.5$ for the REM  ($\square$)
and  of order $1/\nu' =1$ for the DPCT ($\bigcirc$)   }
\label{y2avremcayley}
\end{figure}

As already mentioned in Eq. \ref{y2av}, for $k=2$, the explicit expression
of $\overline{Y_2}(N)$ is particularly simple 
in the low-temperature phase,
\begin{eqnarray}
\overline{ Y_2(T<T_c) }=  \frac{T_c-T}{T_c}
\label{y2avbis}
\end{eqnarray}
In the REM where the only finite-size scaling exponent is $\nu=2$
(Eq. \ref{nu}),
one thus expects at criticality
\begin{eqnarray}
\left[\overline{ Y_2(T_c,N) } \right]_{REM} \oppropto_{N \to \infty} \frac{1}{ N^{1/2} }
\label{y2avremtc}
\end{eqnarray}
in agreement with our numerical simulations, as shown on Fig. \ref{y2avremcayley}.
For the DPCT however, we find numerically that the decay of $\overline{ Y_2}$
is governed by the exponent $\nu'=1$ (Eq. \ref{nuprime}) at criticality
\begin{eqnarray}
\left[ \overline{ Y_2(T_c,N) } \right]_{DPCT} \oppropto_{N \to \infty} \frac{1}{ N }
\label{y2avcayleytc}
\end{eqnarray}
 as shown on Fig. \ref{y2avremcayley}.

\subsection{ Other values of $k$ }

For arbitrary $k$, the explicit value (Eq. \ref{ykrem})
can be expanded in $(T_c-T)/T_c$ as follows
\begin{equation}
\overline{Y_k}(T<T_c)= 
\  \frac{(T_c-T)}{(k-1)T_c} \left[1+ O\left( (T_c-T) \right)\right]
\label{ykremdev}
\end{equation}
For the REM where the only finite-size scaling exponent is $\nu=2$
(Eq. \ref{nu}),
one thus expects at criticality
\begin{eqnarray}
\left[\overline{ Y_k(T_c,N) } \right]_{REM} = \frac{1}{ (k-1) N^{1/2} }
\left[ 1+ O \left( \frac{1}{N^{1/2}} \right) \right]
\label{ykavremtc}
\end{eqnarray}
in agreement with our numerical simulations.
For the DPCT, we find numerically that it is the
 exponent $\nu'=1$ (Eq. \ref{nuprime}) that governs the critical behavior
\begin{eqnarray}
\left[\overline{ Y_k(T_c,N) } \right]_{DPCT} \oppropto \frac{1}{ N }
\left[ 1+ O \left( \frac{1}{ N} \right) \right]
\label{ykavcayleytc}
\end{eqnarray}

\section{ Decay of typical values
 $ Y_k^{typ}(N) = e^{ \overline{\ln Y_k} }$ at criticality }
 
 \label{yktyp}

From the explicit expression of Eq. \ref{lnyklevyres}
of $\overline{ \ln Y_k}$ in the low-temperature phase with $\mu=T/T_c$,
one obtains the following expansion in $(T_c-T)$
\begin{equation}
\overline{ \ln Y_k}=  k \left[ 1+a_1 (T_c-T)+ O( (T_c-T)^2 \right]\ln (T_c-T)
+ b_0+b_1 (T_c-T)+ O( (T_c-T)^2 )
\label{lnyklevydev}
\end{equation}
For the REM where the only finite-size scaling exponent is $\nu=2$
(Eq. \ref{nu}),
one thus expects at criticality
\begin{eqnarray}
\left[\overline{ \ln Y_k(T_c,N) } \right]_{REM} \oppropto 
- \frac{k}{2} \left[ 1+ O\left( \frac{1}{\sqrt N} \right)\right] \ln N 
+ cst + O\left( \frac{1}{\sqrt N} \right)
\label{alnykremtc}
\end{eqnarray}
in agreement with our numerical simulations.

For the DPCT, we find that it is the exponent $\nu'=1$
that governs the decay of typical weights 
\begin{eqnarray}
\left[\overline{ \ln Y_k(T_c,N) } \right]_{DPCT} \oppropto 
- k \left[ 1+ O\left( \frac{1}{N} \right)\right] \ln N 
+ cst + O\left( \frac{1}{N} \right)
\label{alnykcayleytc}
\end{eqnarray}

To summarize, the exponents governing the decay of typical values
are exactly linear in $k$ in both models
\begin{eqnarray}
\left[ Y_k^{typ}(N) \right]_{REM} = e^{ \overline{\ln Y_k} } 
\propto \frac{1}{N^{k/2}}
\label{yktypremtc} 
\end{eqnarray}
and
\begin{eqnarray}
\left[ Y_k^{typ}(N) \right]_{DPCT} = e^{ \overline{\ln Y_k} } 
\propto \frac{1}{N^{k}}
\label{yktypcayleytc} 
\end{eqnarray}
in contrast with disorder averaged values, where the exponents do not depend on $k$
(Eqs \ref{ykavremtc} and \ref{ykavcayleytc} ).
To explain this difference, we now discuss the histograms of 
$w_{max}$ and of $Y_2$ at criticality.

\section{ Probability distributions of $w_{max}$ and of $Y_2$ at criticality }

\label{histo}

\subsection{  Probability distribution of $w_{max}(N)$ at criticality}

\begin{figure}[htbp]
\includegraphics[height=6cm]{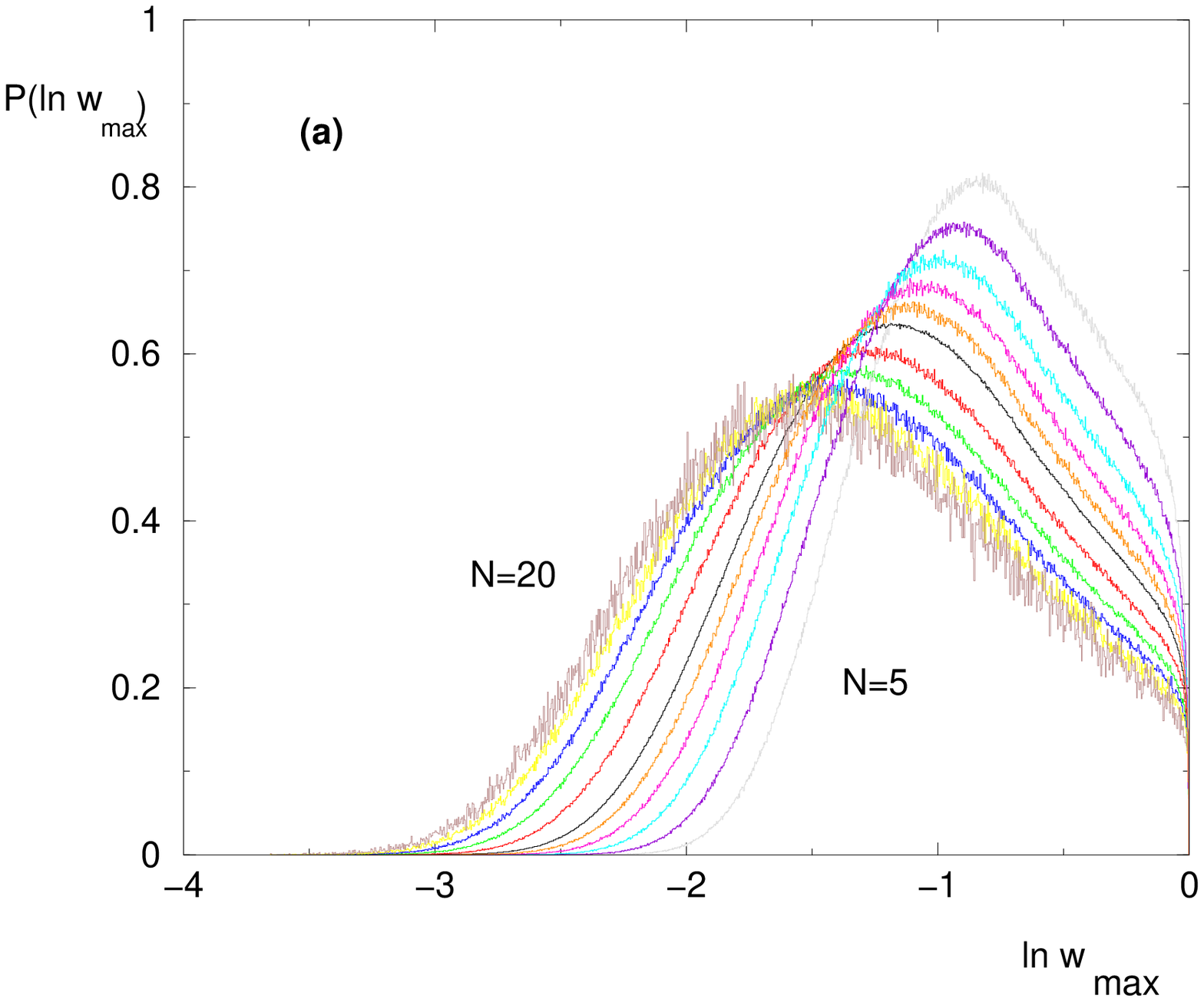}
\hspace{1cm}
\includegraphics[height=6cm]{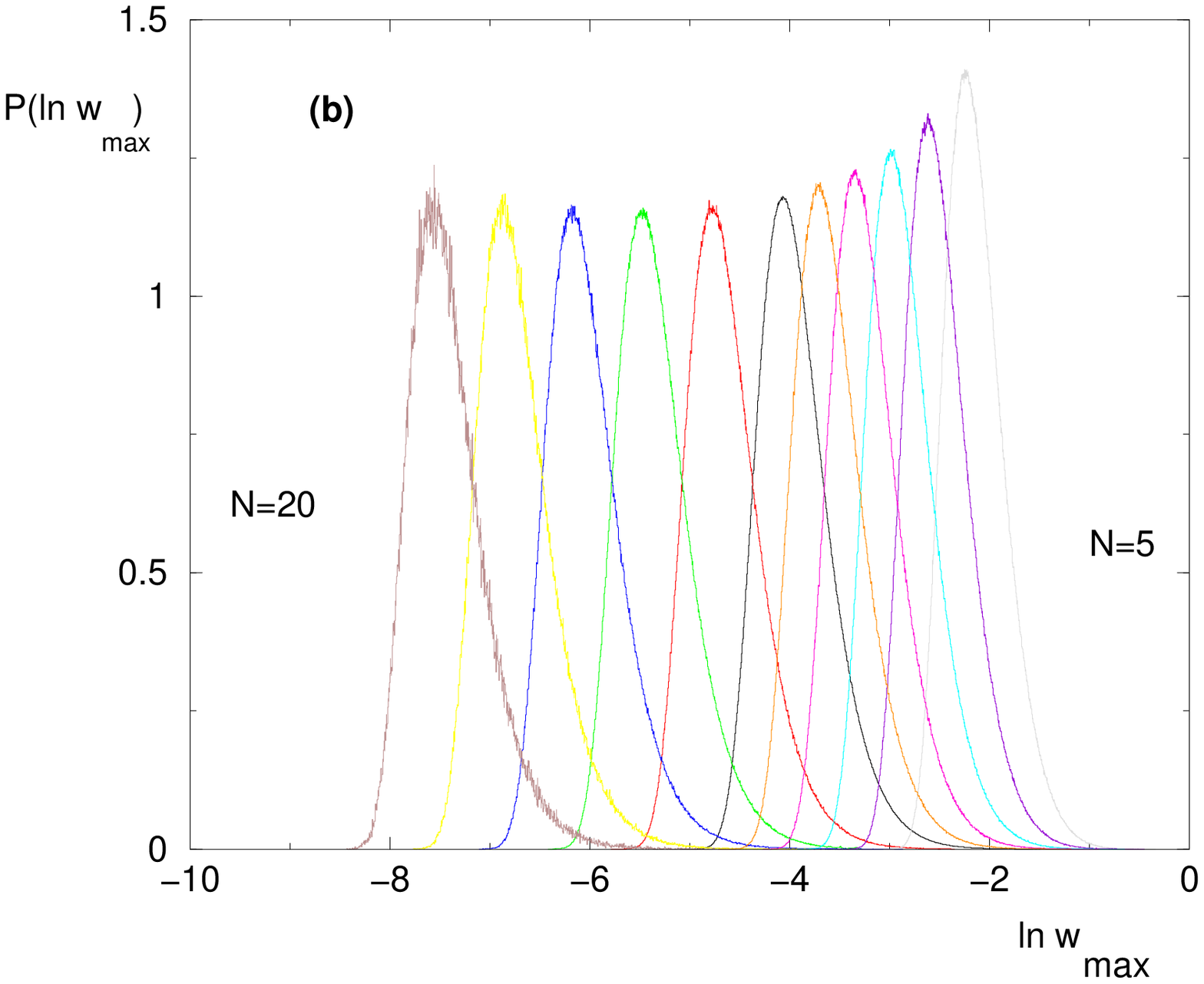}
\caption{ (Color on line) REM :  Probability distribution $P_N(\ln w_{max})$ of the
maximal weight among the $2^N$ configurations,
for sizes $N=5,6,7,8,9,10,12,14,16,18,20$
(a)  at $T_c$ :  the distribution remains broad around the
averaged value $\overline{\ln w_{max}} \simeq - (\ln N)/2$, with 
a slow decay of rare events near the origin $\ln w_{max} \sim 0$.
 (b)  at $T=2>T_c$ :   the width around the average value
$\overline{\ln w_{max} } \propto - N$ converges towards a constant. }
\label{wmaxremhisto}
\end{figure}

\begin{figure}[htbp]
\includegraphics[height=6cm]{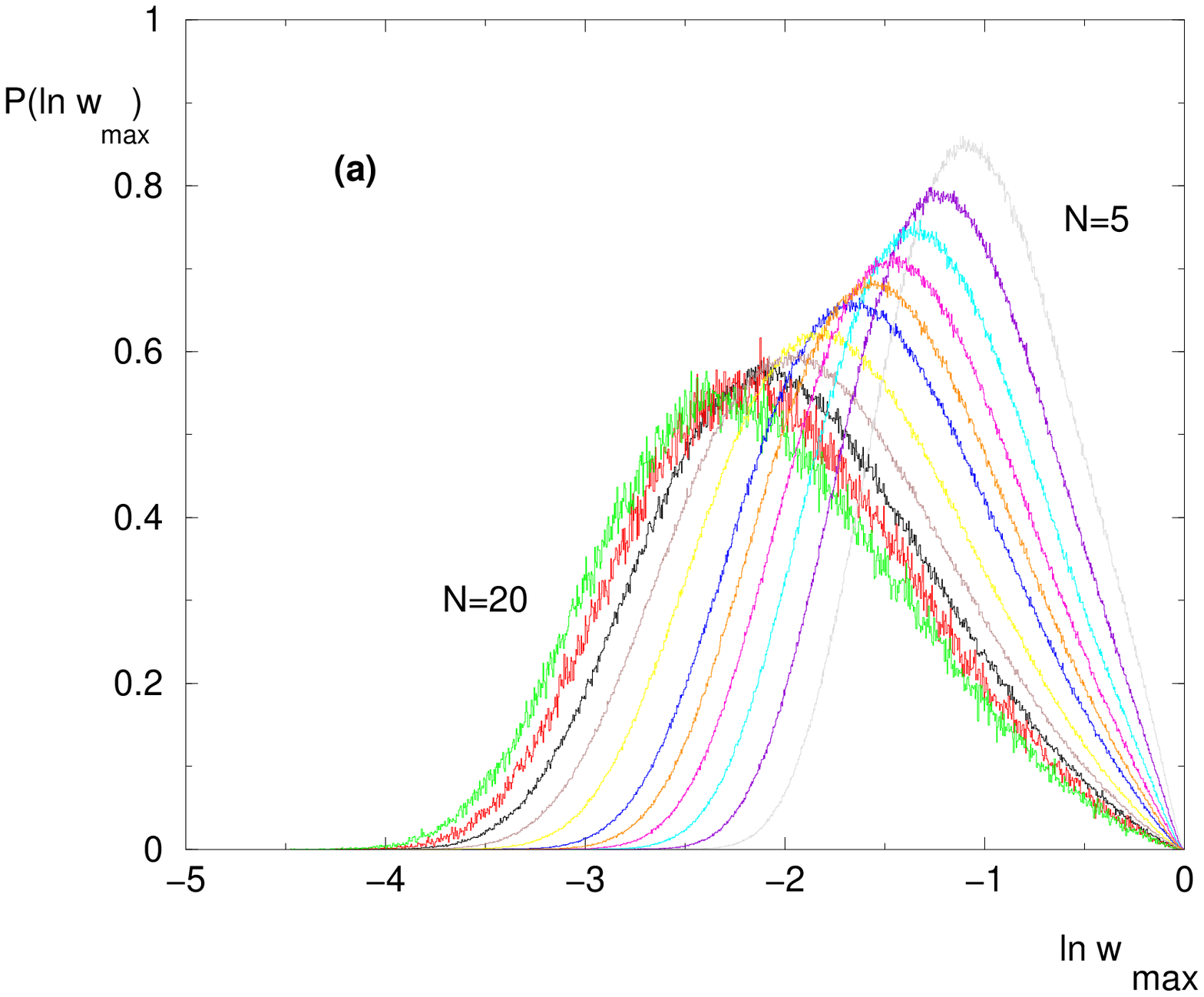}
\hspace{1cm}
\includegraphics[height=6cm]{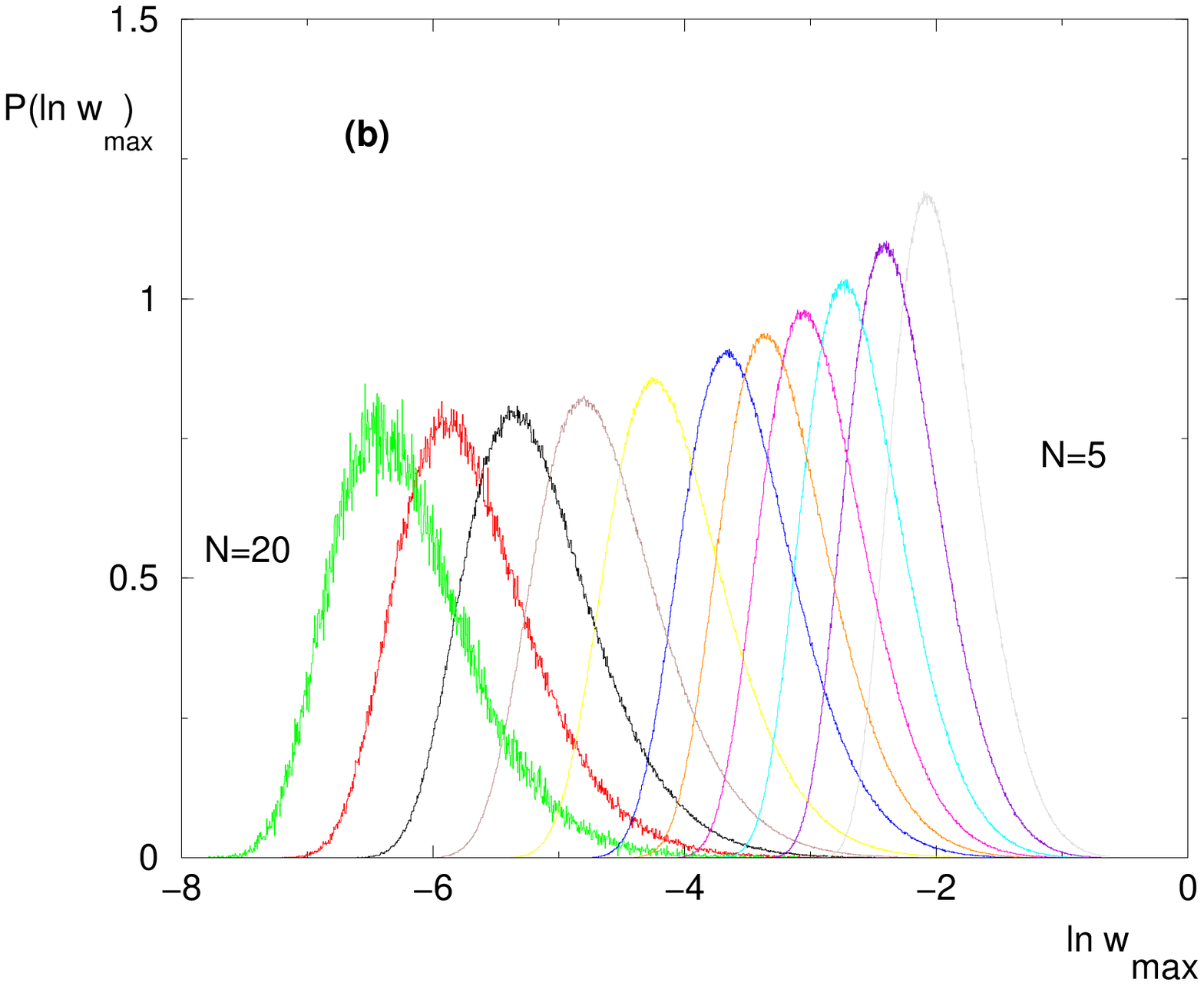}
\caption{ (Color on line) DPCT : Probability distribution $P_N(\ln w_{max})$ of the
maximal weight among the $2^N$ configurations,
for sizes $N=5,6,7,8,9,10,12,14,16,18,20$
(a)  at $T_c$ : the distribution remains broad around the
averaged value $\overline{\ln w_{max}} \simeq - (\ln N)$, with 
a slow decay of rare events near the origin $\ln w_{max} \sim 0$.
 (b)  at $T=2>T_c$ :  the width around the average value
$\overline{\ln w_{max} } \propto -N$ converges towards a constant. }
\label{wmaxcayleyhisto}
\end{figure}

For each sample, we consider the maximal weight 
\begin{eqnarray}
w_{max}  = max_{i} \left[ w_i \right]
\end{eqnarray}
among the $2^N$ configurations. We show on Fig. \ref{wmaxremhisto}
and \ref{wmaxcayleyhisto} the probability distribution $P_N( \ln w_{max})$
over the samples for the REM and the DPCT, both at criticality
and in the high temperature phase for comparison.
At criticality, $P_N(\ln w_{max})$ remains broad around the
averaged value 
\begin{eqnarray}
\overline{\ln w_{max}} &&  \simeq - \frac{1}{2} (\ln N)+...
 {\rm \ \ for \ the \ REM } \\
\overline{\ln w_{max}} &&  \simeq - (\ln N)+... 
{\rm \ \ for \ the \ DPCT } 
\label{wmaxtyptc} 
\end{eqnarray}
 with a slow decay of rare events near the origin $\ln w_{max} \sim 0$.
Again, as for the entropy distribution (see Figs \ref{sremhisto} (a)
 and \ref{scayleyhisto} (a) ), the statistics of these rare 'still frozen' samples
is not the same in the REM and in the DPCT as shown on Figs. \ref{wmaxremhisto} (a)
and \ref{wmaxcayleyhisto} (a).
In the high temperature phase, the width around the average value
$\overline{\ln w_{max} } \propto - N$ converges towards a constant
in both models, as shown on Fig.  \ref{wmaxremhisto} (b)
and \ref{wmaxcayleyhisto} (b).

\newpage

\subsection{ Probability distribution of $Y_2$ at criticality}

\begin{figure}[htbp]
\includegraphics[height=6cm]{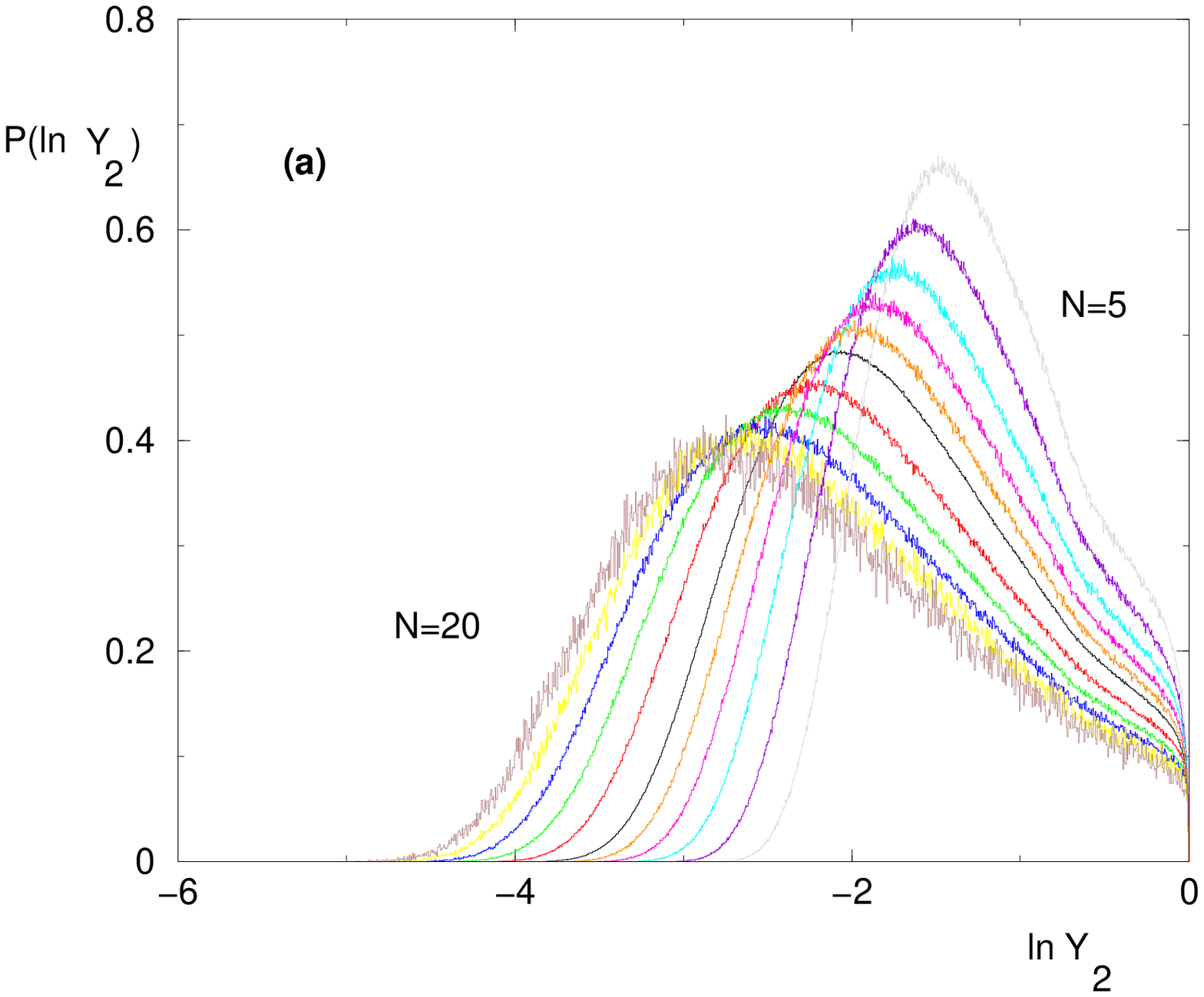}
\hspace{1cm}
\includegraphics[height=6cm]{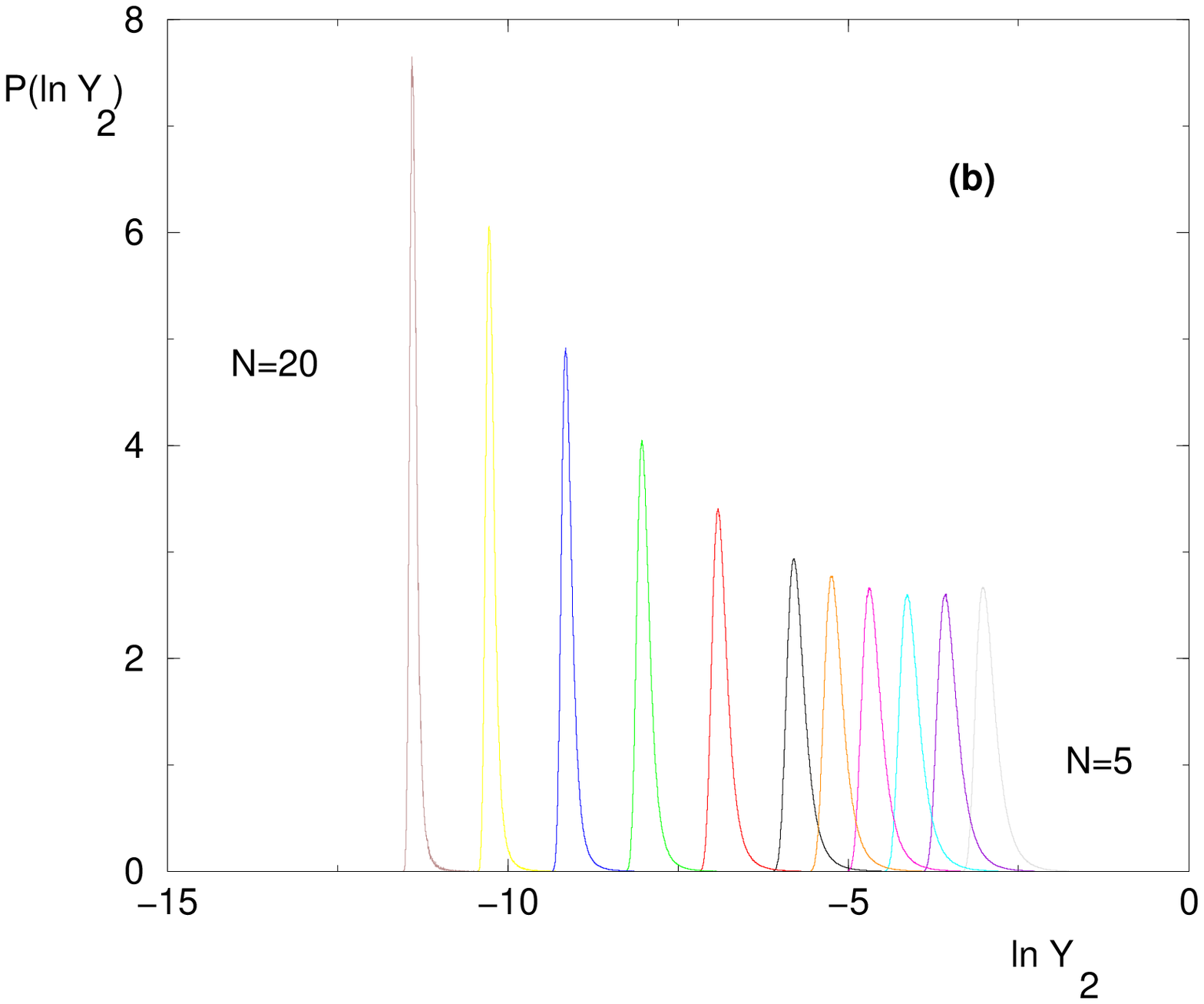}
\caption{ (Color on line) REM : Probability distribution $P_N(\ln Y_2)$ 
for sizes $N=5,6,7,8,9,10,12,14,16,18,20$
(a)  at $T_c$ :  the distribution remains broad around the
averaged value $\overline{\ln Y_2} \simeq - (\ln N)$, with 
a slow decay of rare events near the origin $\ln Y_2 \sim 0$.
 (b)  at $T=2>T_c$ : the width around the average value
$\overline{\ln Y_2 } \propto - N$ converges towards zero.  }
\label{y2remhisto}
\end{figure}

\begin{figure}[htbp]
\includegraphics[height=6cm]{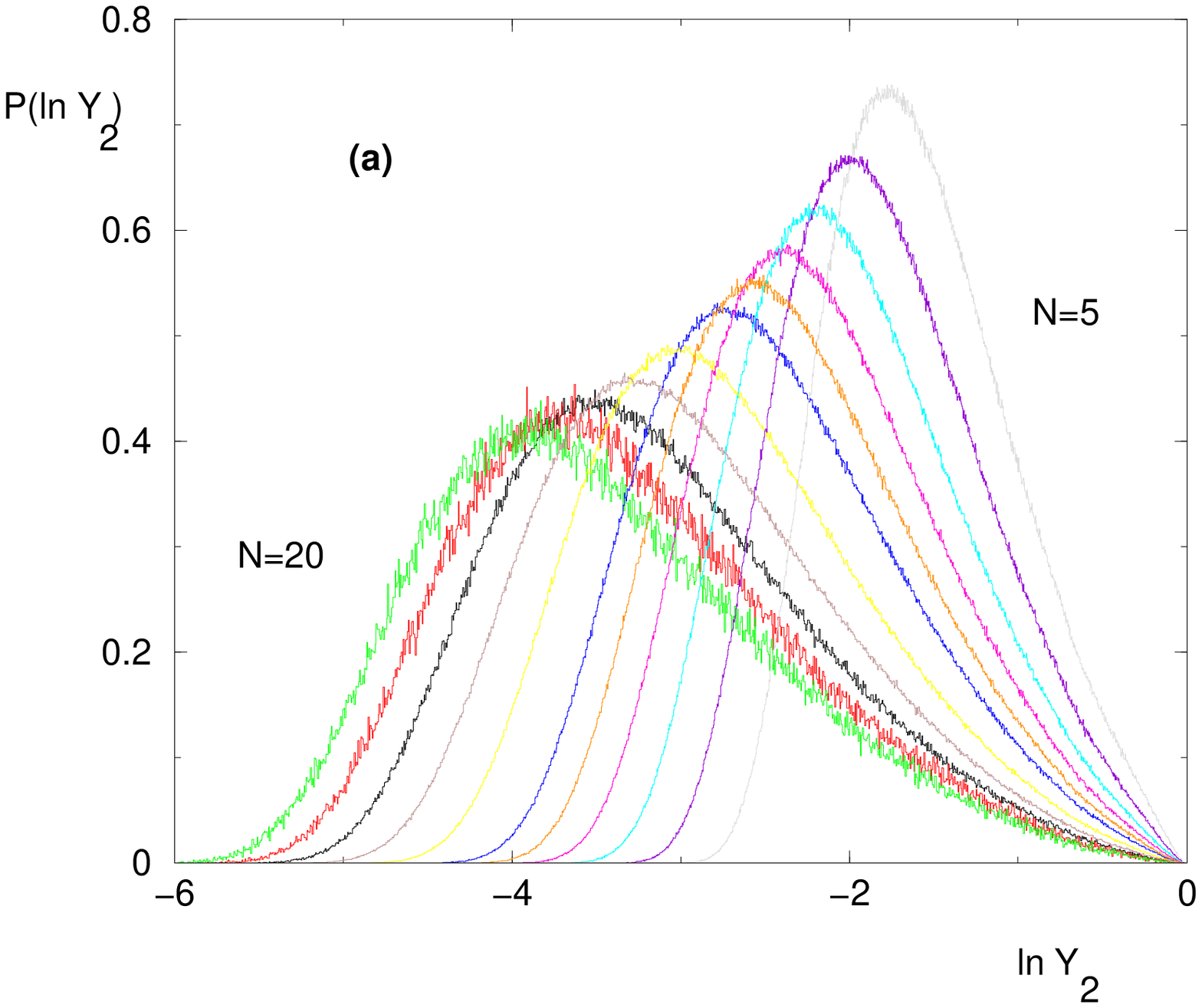}
\hspace{1cm}
\includegraphics[height=6cm]{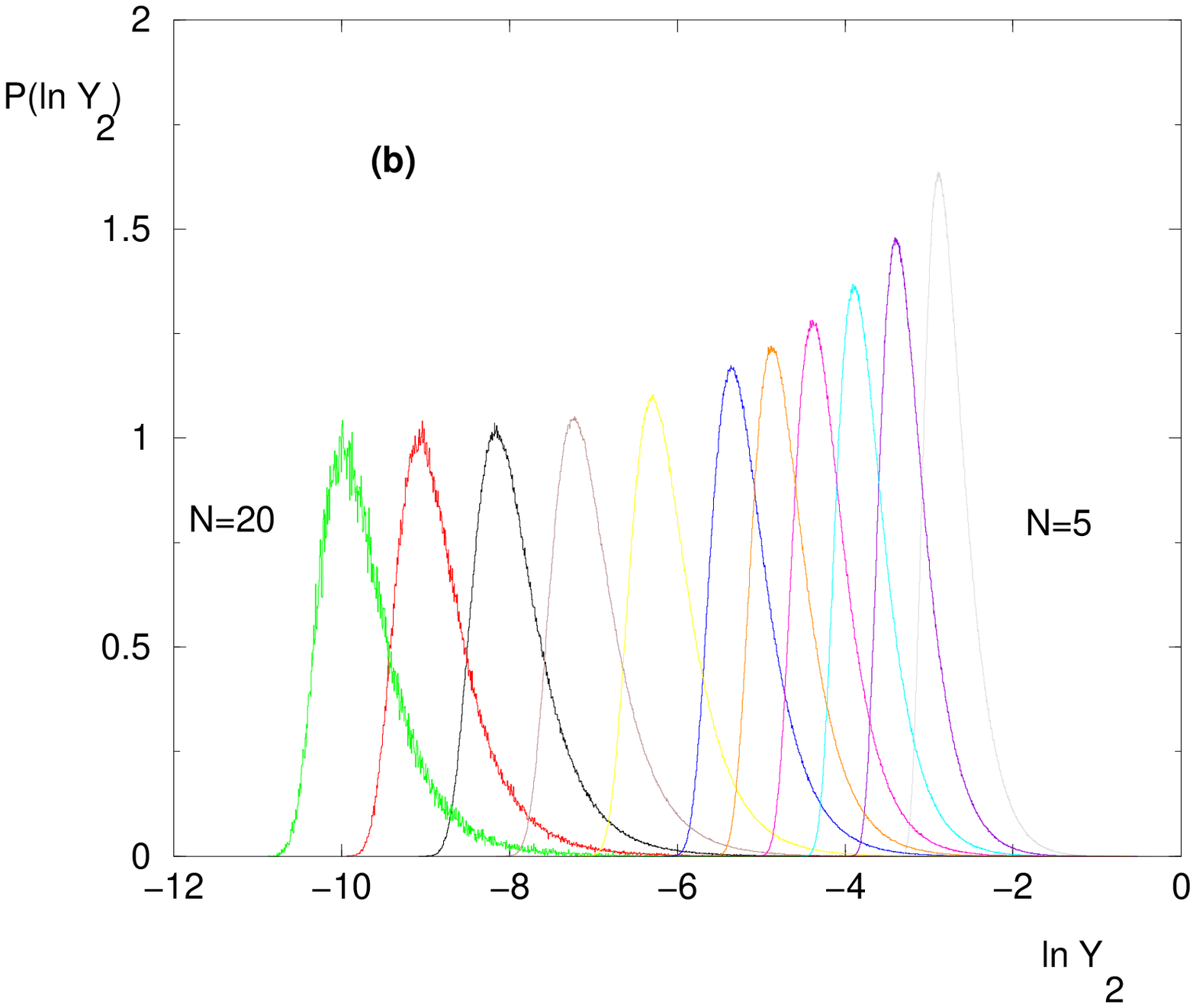}
\caption{ (Color on line) DPCT :  Probability distribution $P_N(\ln Y_2)$ 
for sizes $N=5,6,7,8,9,10,12,14,16,18,20$
(a)  at $T_c$ : the distribution remains broad around the
averaged value $\overline{\ln Y_2} \simeq - 2 (\ln N)$, with 
a slow decay of rare events near the origin $\ln Y_2 \sim 0$.
 (b) at $T=2>T_c$ : the width around the average value
$\overline{\ln Y_2 } \propto N$ converges towards a constant.  }
\label{y2cayleyhisto}
\end{figure}

 We show on Fig. \ref{y2remhisto}
and \ref{y2cayleyhisto} the probability distribution $P_N( \ln Y_2)$
over the samples for the REM and the DPCT, both at criticality
and in the high temperature phase for comparison.
At criticality, $P_N(\ln Y_2)$ remains broad around the
averaged value 
\begin{eqnarray}
\overline{\ln Y_2} &&  \simeq -  (\ln N)+... {\rm \ \ for \ the \ REM } \\
\overline{\ln Y_2} &&  \simeq - 2 (\ln N)+... {\rm \ \ for \  the \ DPCT } \\
\label{y2typtc} 
\end{eqnarray}
 with again a different decay of rare events near the origin $\ln Y_2 \sim 0$.
In the high temperature phase, the width around the average value
$\overline{\ln Y_2 } \propto - N$ converges towards 
zero for the REM, as shown on Fig.  \ref{y2remhisto} (b),
and towards a finite constant for the DPCT 
 as shown on Fig.  \ref{y2cayleyhisto} (b).

As a final remark, it is interesting to compare the  probability distribution of $\ln Y_2$
at criticality for the directed polymer on the Cayley tree (Fig. \ref{y2cayleyhisto} a)
and in dimension $1+3$ (see Fig. 3 a of \cite{DP3dmultif}), where as $N$ grows, 
the distribution simply shifts along the $x$-axis with a fixed shape.

\subsection{ Rare events where $w_{max} \sim 1$ and $Y_2 \sim 1$ }

\label{rare}

\begin{figure}[htbp]
\includegraphics[height=6cm]{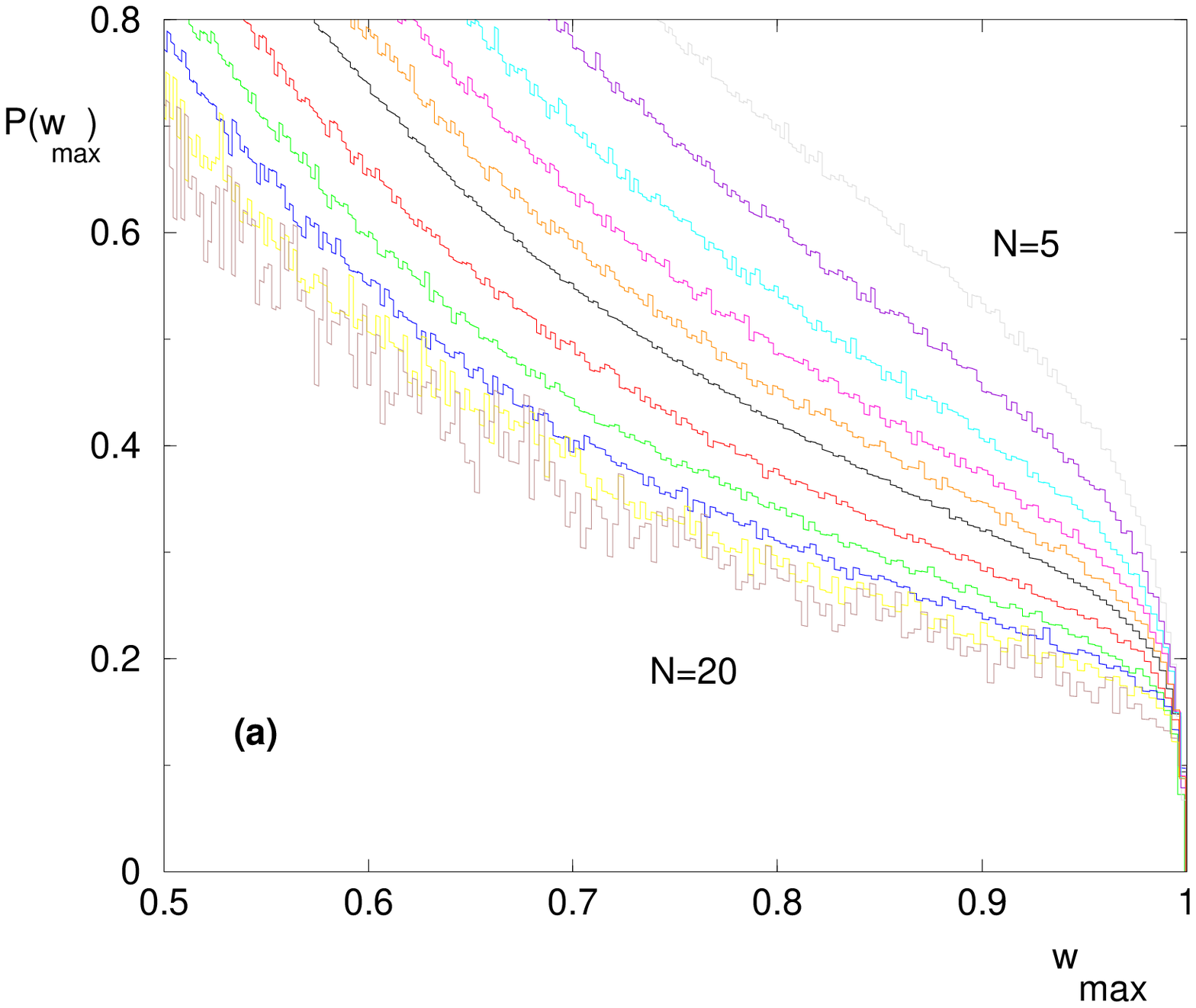}
\hspace{1cm}
 \includegraphics[height=6cm]{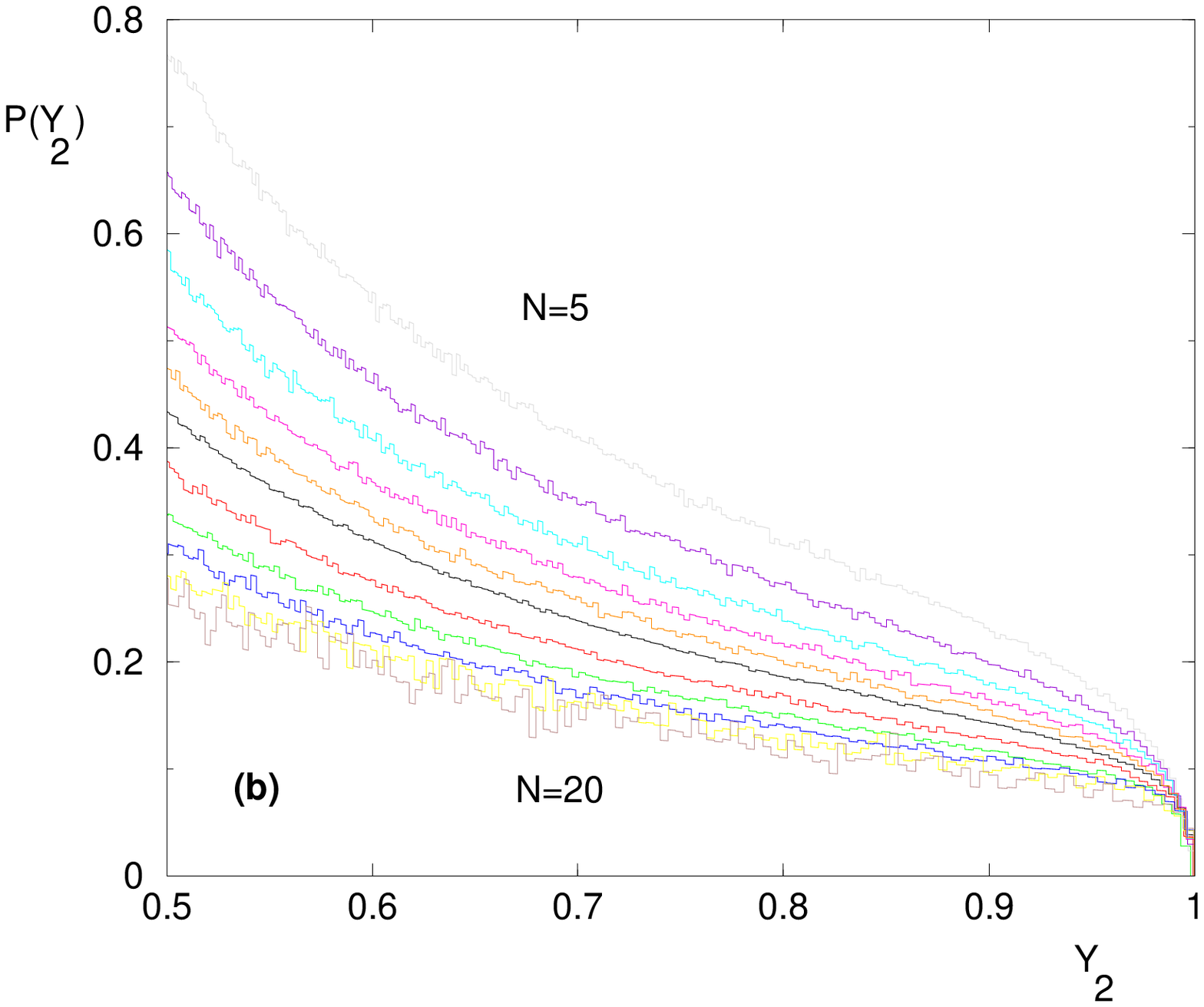}
\caption{ (Color on line) REM at criticality : statistics of
rare finite samples which are still 'frozen' at $T_c$
for sizes $N=5,6,7,8,9,10,12,14,16,18,20$
(a) Probability distribution of the maximal weight
 $w_{max}$ in the region $1/2 \leq w_{max} \leq 1$
 (b) Probability distribution of $Y_2$ 
in the region $1/2 \leq Y_2 \leq 1$  }
\label{rareremhisto}
\end{figure}

\begin{figure}[htbp]
\includegraphics[height=6cm]{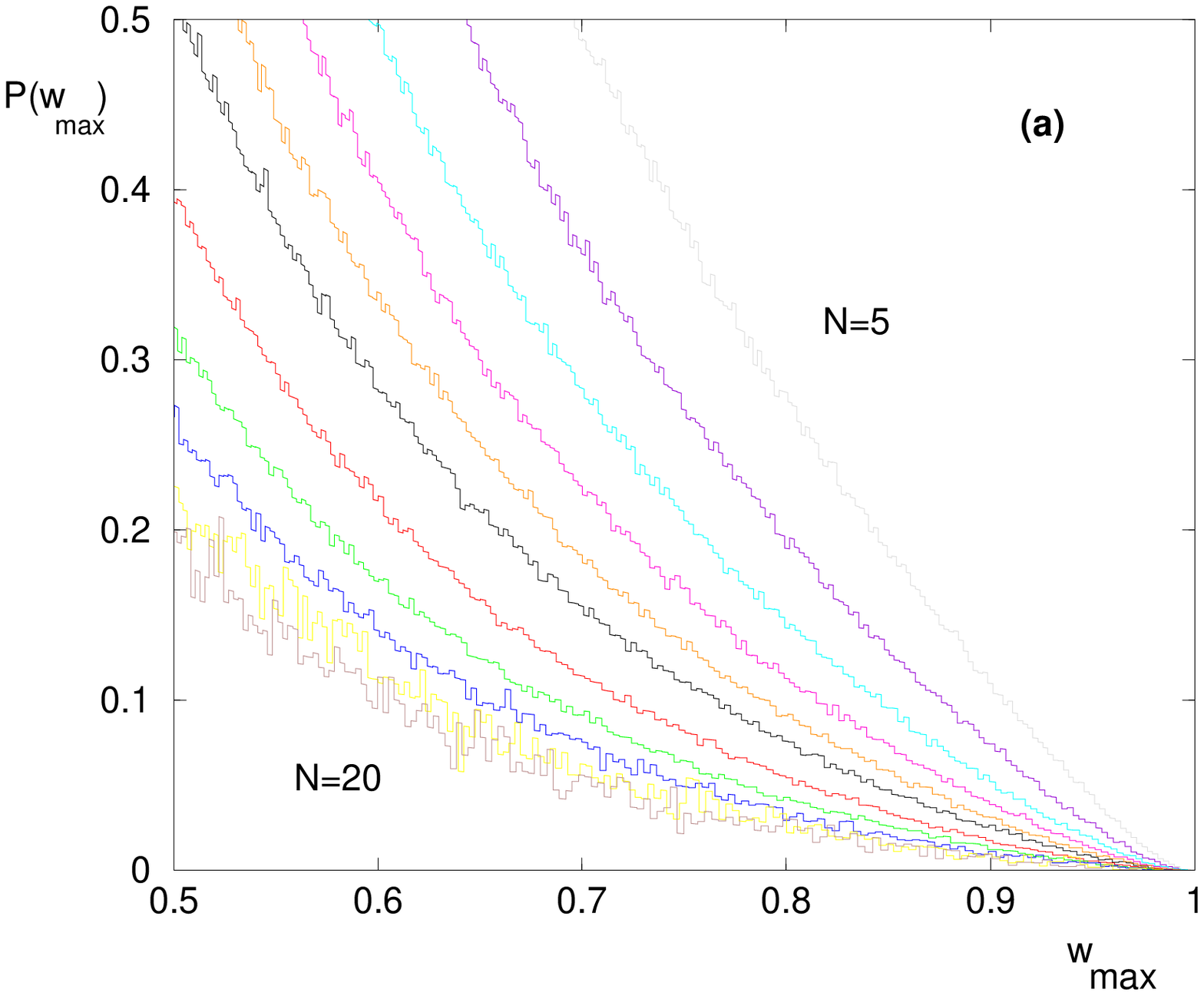}
\hspace{1cm}
\includegraphics[height=6cm]{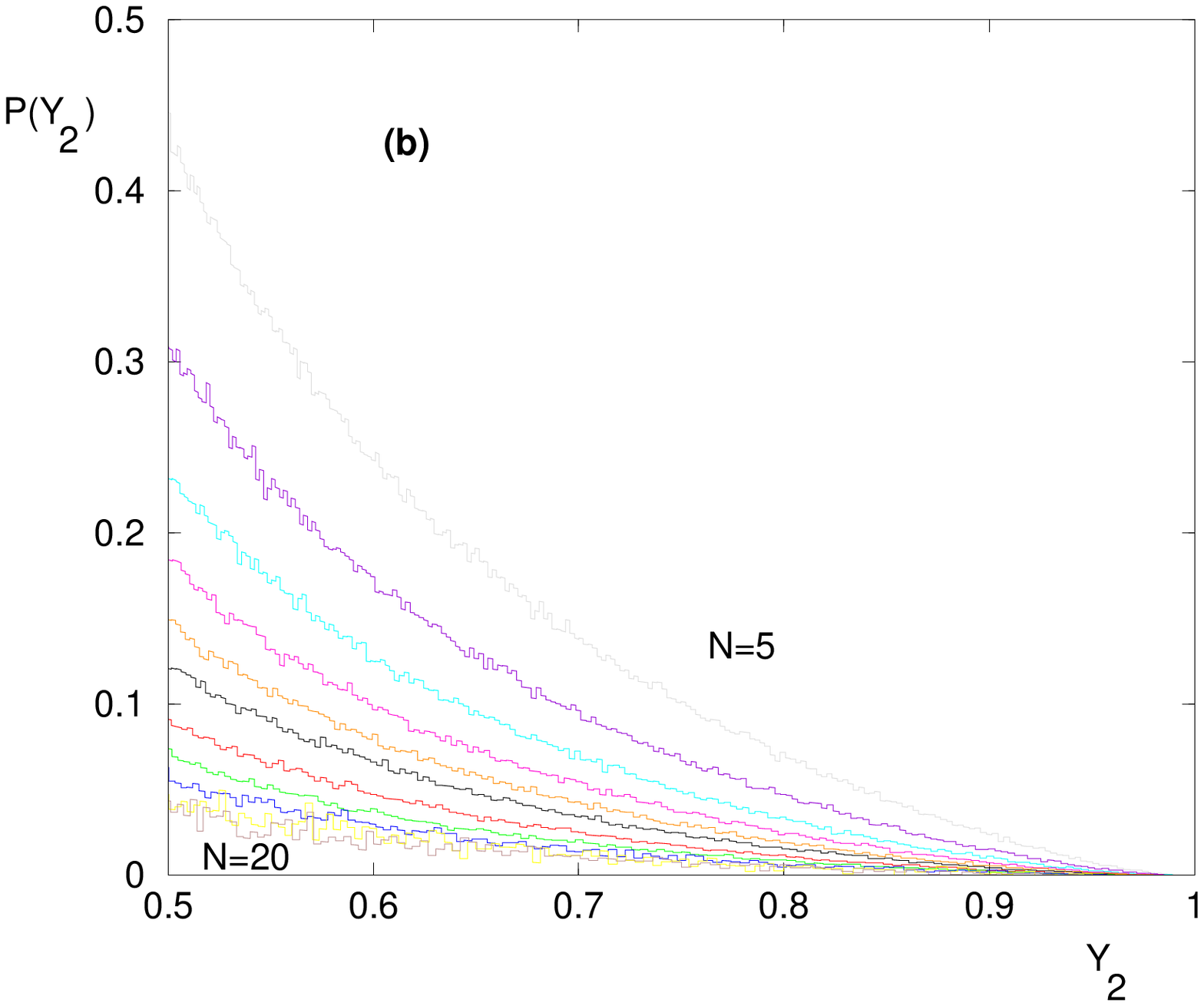}
\caption{ (Color on line) DPCT at criticality : statistics of
rare finite samples which are still 'frozen' at $T_c$
for sizes $N=5,6,7,8,9,10,12,14,16,18,20$
(a) Probability distribution of the maximal weight for sizes
 $w_{max}$ in the region $1/2 \leq w_{max} \leq 1$ :
note the difference with the REM (Fig. \ref{rareremhisto} a  )
 (b) Probability distribution of $Y_2$ 
in the region $1/2 \leq Y_2 \leq 1$ :
 note the difference with the REM (Fig. \ref{rareremhisto} b) }
\label{rarecayleyhisto}
\end{figure}

In the low temperature phase $T<T_c$, the statistical properties of
$w_{max}$ and $Y_2$ have been studied in details in \cite{Der_Fly}.
In particular, the probability distribution $P_{T<T_c}(w_{max})$
coincides for $1/2<w_{max} \leq 1$ with the weight density $f(w)$
given in Eq. \ref{densitew}. Near the transition $\mu=T/T_c \to 1^-$,
the singularity near $w_{max}$ reads 
\begin{equation}
P_{T<T_c}(w_{max}) = f_{T<T_c}(w_{max})
\opsimeq_{w_{max} \to 1} (1- \mu ) (1-w_{max})^{\mu-1}
\label{p1wlow}
\end{equation}
In this section, we are interested in the behavior of
 the probability distribution $P_{T_c,N}(w_{max})$ near $w_{max} \to 1$
for finite samples at criticality
\begin{eqnarray}
P_{T_c,N}(w_{max}) = f_{T_c,N}(w_{max}) \opsimeq_{w_{max} \to 1} A_N (1-w_{max})^{\sigma}
\label{ptcnwmax}
\end{eqnarray}
The same singularity governs the probability distribution of $Y_2$
\begin{eqnarray}
P_{T_c,N}(Y_2)  \opsimeq_{ Y_2 \to 1} A_N (1- Y_2)^{\sigma}
\label{ptcny2}
\end{eqnarray}

The amplitude $A_N$ represents the global scaling of the rare samples
which are 'still frozen', whereas the exponent $\sigma$ describes
the shape of the singularity.
These rare events govern the disorder-averaged values $\overline{Y_k}$
at criticality (Eq. \ref{lienykf}), and for large $k$, 
the exponent $\sigma$ governs the power-law dependence in $k$
\begin{eqnarray}
\overline{Y_k}(N) \oppropto_{ k \to \infty} \frac{A_N}{ k^{1+\sigma}}
\label{ykmoycayleykgrand} 
\end{eqnarray}

For the REM, where all finite-size scaling properties involve the
factor $(T_c-T) N^{1/2}$, we expect
\begin{eqnarray}
A_N^{REM} && \oppropto_{N \to \infty} \frac{1}{N^{1/2}}  \\
\sigma_{REM} && =0 
\label{sigmarem}
\end{eqnarray}
This is in agreement via Eq. \ref{ykmoycayleykgrand} 
 with the leading behavior of the disorder-averaged values $\overline{Y_k}$
of Eq. \ref{ykavremtc}.
We show on Fig. \ref{rareremhisto} the behavior of the probability distributions
of $w_{max}$ and $Y_2$ near $w_{max} \to 1$ and $Y_2 \to 1$

For the DPCT, the situation is more subtle.
From the behavior in $N$ of disorder-averaged values of Eq. \ref{ykavcayleytc},
we conclude that the amplitude is governed by $\nu'=1$
\begin{eqnarray}
A_N^{DPCT} \oppropto_{N \to \infty} \frac{1}{N} 
\end{eqnarray}
However in contrast with the REM, 
the behavior of the probability distributions
of $w_{max}$ and $Y_2$ near $w_{max} \to 1$ and $Y_2 \to 1$ 
shown of Fig. \ref{rarecayleyhisto} corresponds to a value $\sigma >1$
for the singularity exponent.
The measure of the $k$-dependence of Eq. \ref{ykmoycayleykgrand}
indeed leads to a value of order
\begin{eqnarray}
\sigma_{DPCT} \sim 1.5
\label{sigmacayley}
\end{eqnarray}

The fact that a finite $\sigma$ appears at criticality for the DPCT,
in contrast with the REM where $\sigma=0$ in continuity
with the low-temperature phase,
indicates that the tree structure plays a role at criticality,
in contrast with the low-temperature phase where the overlap 
distribution is concentrated on $q=0$ and $q=1$ (Eq. \ref{piqlow}).
In the next Section, we describe the finite-size properties
of the overlap distribution at criticality.

\section{ Overlap distribution at criticality }

\label{overlap}

In disorder-dominated phases, the order parameter is the 'overlap' between
two thermal configurations in the same disordered sample.
In this Section, we discuss in detail the overlap distribution for the DPCT,
and compare with the REM case in the end.

For the DPCT, we consider the probability
$P_N(t)$ that two walks of $N$ steps have $t$ common bonds
in a fixed sample of a Cayley tree, where the possible values are $t=0,1,..N$.
The normalization reads
\begin{eqnarray}
\sum_{t=0}^N P_N(t) =1
\label{pnt}
\end{eqnarray}
The usual overlap distribution $\pi_N(q)$ concerning the fraction
$q=t/N$ of common bonds reads 
\begin{eqnarray}
\pi_N(q)= N  P_N(t=Nq )
\label{defpiq}
\end{eqnarray}
with the normalization
\begin{eqnarray}
\int_0^1 dq \pi_N(q)= 1
\label{normapiq}
\end{eqnarray}

\subsection{ Reminder on the overlap distribution for $T<T_c$ for the DPCT  }

As recalled in Eq. \ref{piqlow}, 
the distribution of the overlap $q$ between two walks
on the same disordered tree is simply the sum of
two delta peaks at $q=0$ and $q=1$ in
the whole low-temperature phase \cite{Der_Spo},
and in particular the disorder average over the samples reads (Eq. \ref{y2av})
\begin{eqnarray}
{ \overline \pi}_{N=\infty} (q)= \frac{T}{T_c} \delta(q) +
\left( 1- \frac{T}{T_c} \right)\delta(q-1)
\label{piqlowav}
\end{eqnarray}
The finite-size corrections have been studied in \cite{Fis_Hus,Tan} :
the probability $P_N(t)$ is finite at $t=0$ and at $t=N$,
whereas for $0 \ll t \ll N$, 
the disorder averaged probability $P_N(t)$ obeys the
scaling 
\begin{eqnarray}
{\overline P}_N( 0<t<N)  \oppropto_{N \to \infty} \frac{1}{N^{3/2}}
\psi \left( \frac{t}{N} \right)
\label{scalingpntlow}
\end{eqnarray}
where the function $\psi(q)$ presents the singularities $q^{-3/2}$
and $(1-q)^{-3/2}$ near the two boundaries $q \to 0$ and $q \to 1$.
For the finite-size overlap distribution, 
Eq. \ref{scalingpntlow} translates into the finite size correction
\begin{eqnarray}
{ \overline \pi}_N(0<q<1) \oppropto_{N \to \infty} \frac{1}{N^{1/2}} \psi \left( q \right)
\end{eqnarray}
to the asymptotic result of Eq. \ref{piqlowav}.

\subsection{ Finite-size overlap distribution at $T_c$ for the DPCT  }

\begin{figure}[htbp]
\includegraphics[height=6cm]{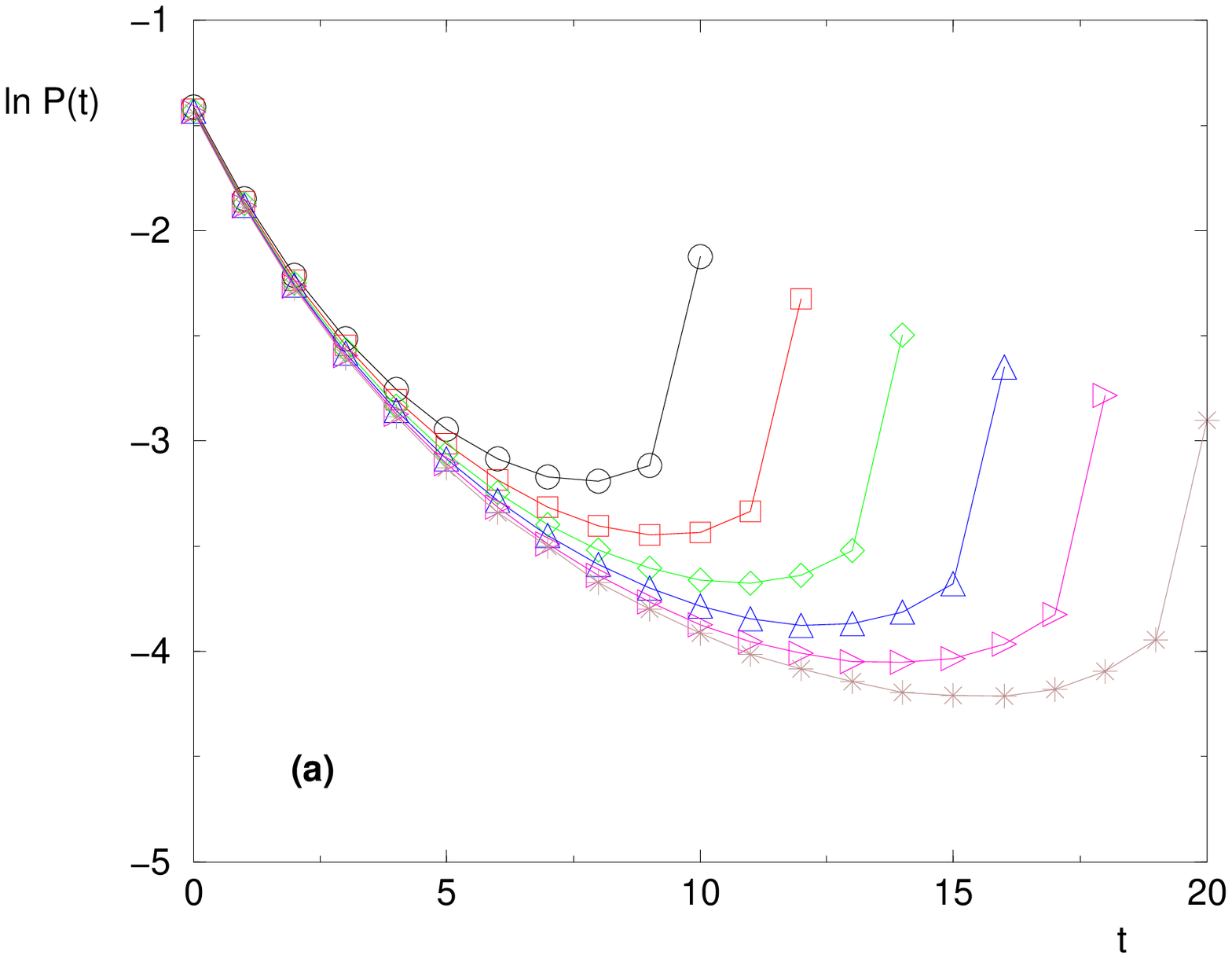}
\hspace{1cm}
 \includegraphics[height=6cm]{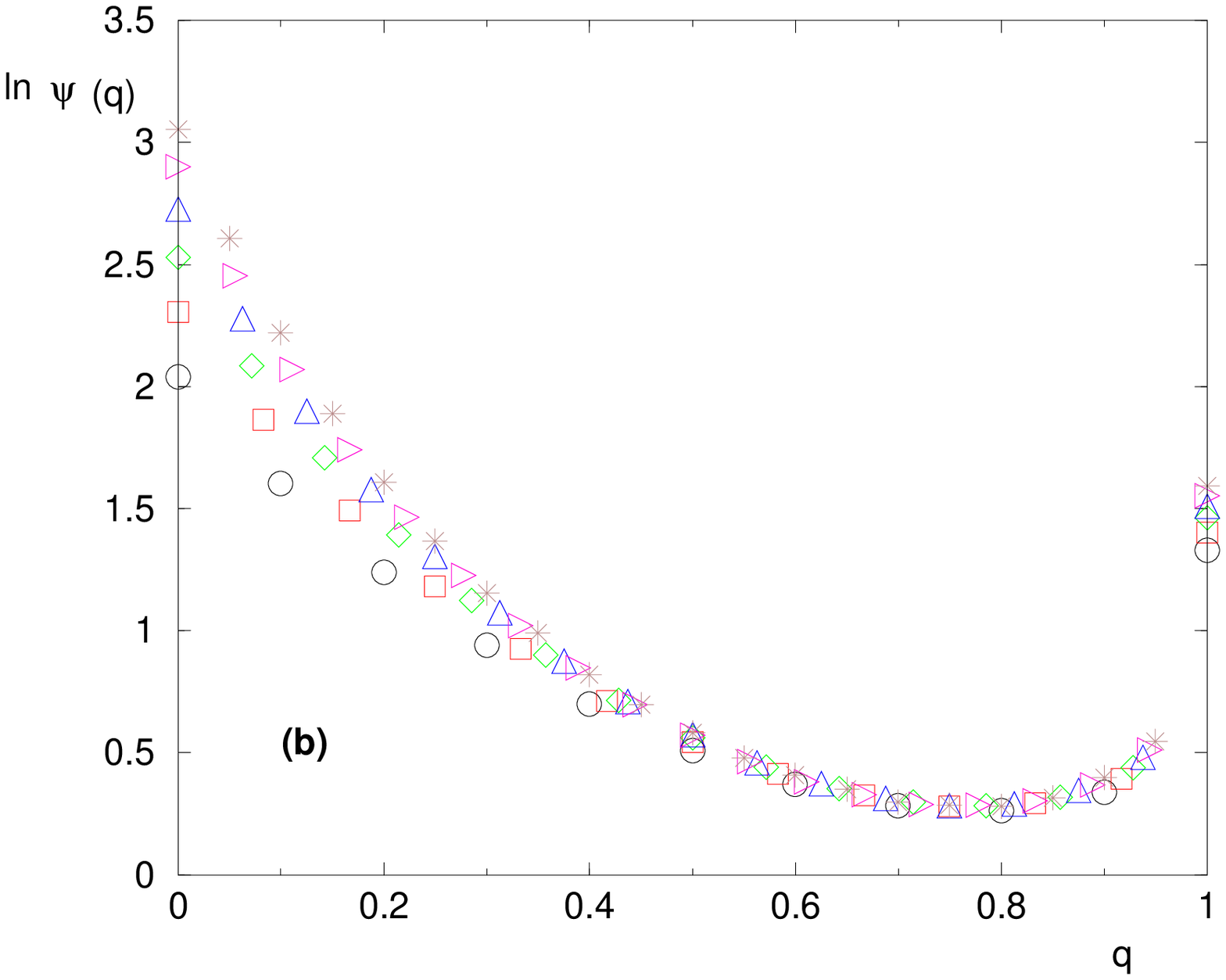}
\caption{ (Color on line) DPCT at criticality  
(a) Logarithm of the disorder averaged probability distribution $\overline{P}_N(t)$
of the number $t=0,1,..N$ of common bonds between two walks
for sizes $N=10 (\bigcirc), 12 (\square), 14 (\lozenge ), 16 (\triangle), 18 (\rhd),
20 (\ast)$
 (b) test of the scaling form of Eq. \ref{scalingpinq} 
for the disorder averaged probability distribution $\overline{\pi}_N(q)$ of the overlap
 $ 0 \leq q=t/N \leq 1$ : $\ln \psi(q)=\ln (N^{0.5}\overline{\pi}_N(q) )$ as a function of $q$}
\label{overlaptc}
\end{figure}

In the limit $N \to \infty$, Eq. \ref{piqlowav} becomes for $T=T_c$
\begin{eqnarray}
{\overline \pi}_{T_c,N=\infty} (q) = \delta(q) 
\label{piqtcav}
\end{eqnarray}
i.e. the whole normalization is concentrated on $q=0$.
Here we are interested into the finite-size corrections to this result.
We show on Figs \ref{overlaptc} (a) and (b) the probability
distributions ${ \overline P}_{T_c,N}(t)$ and ${\overline \pi}_{T_c,N}(q)$
for various sizes. We now discuss the intermediate region $0<q<1$
and the two limit values $q=0$ and $q=1$.

\subsubsection{ Region of intermediate overlap $0<q<1$}

For $0<t<1$, we find numerically that 
 the disorder averaged probability ${ \overline P}_N(t)$ obeys the
scaling 
\begin{eqnarray}
{ \overline P}_N( 0 \ll t \ll N)  \oppropto_{N \to \infty} \frac{1}{N^{1.5}}
\psi \left( \frac{t}{N} \right)
\label{scalingpnt}
\end{eqnarray}
or equivalently for the disorder averaged overlap distribution ${ \overline \pi}_N(q)$ of Eq. \ref{defpiq}
\begin{eqnarray}
{ \overline \pi}_N(0<q<1)  \oppropto_{N \to \infty} \frac{1}{N^{0.5}}
\psi \left( q \right)
\label{scalingpinq}
\end{eqnarray}
as shown on Fig. \ref{overlaptc} b.

\subsubsection{ Region of zero overlap $q=0$ at criticality}

For finite $t$ and $N \to \infty$, 
${ \overline P}_N(t)$ converge to finite values
as shown on Fig. \ref{overlaptc} a, in particular
\begin{eqnarray}
{ \overline P}_N(t=0 )  \opsimeq_{N \to \infty} 0.23 \\
{ \overline P}_N(t=1 )  \opsimeq_{N \to \infty} 0.15
\label{smallt}
\end{eqnarray}
such that the normalization of these finite values corresponding to $q=0$
after rescaling, is $1$ (Eq. \ref{piqtcav}).
From the matching with the scaling regime of Eq. \ref{scalingpnt},
one expects the following power-law decay for large $t$
\begin{eqnarray}
{ \overline P}_{N=\infty}( t )  \oppropto_{t \to \infty} \frac{1}{t^{1.5}}
\label{pninfty}
\end{eqnarray}

\subsubsection{ Probability of full overlap $q=1$ at criticality}

\begin{figure}[htbp]
\includegraphics[height=6cm]{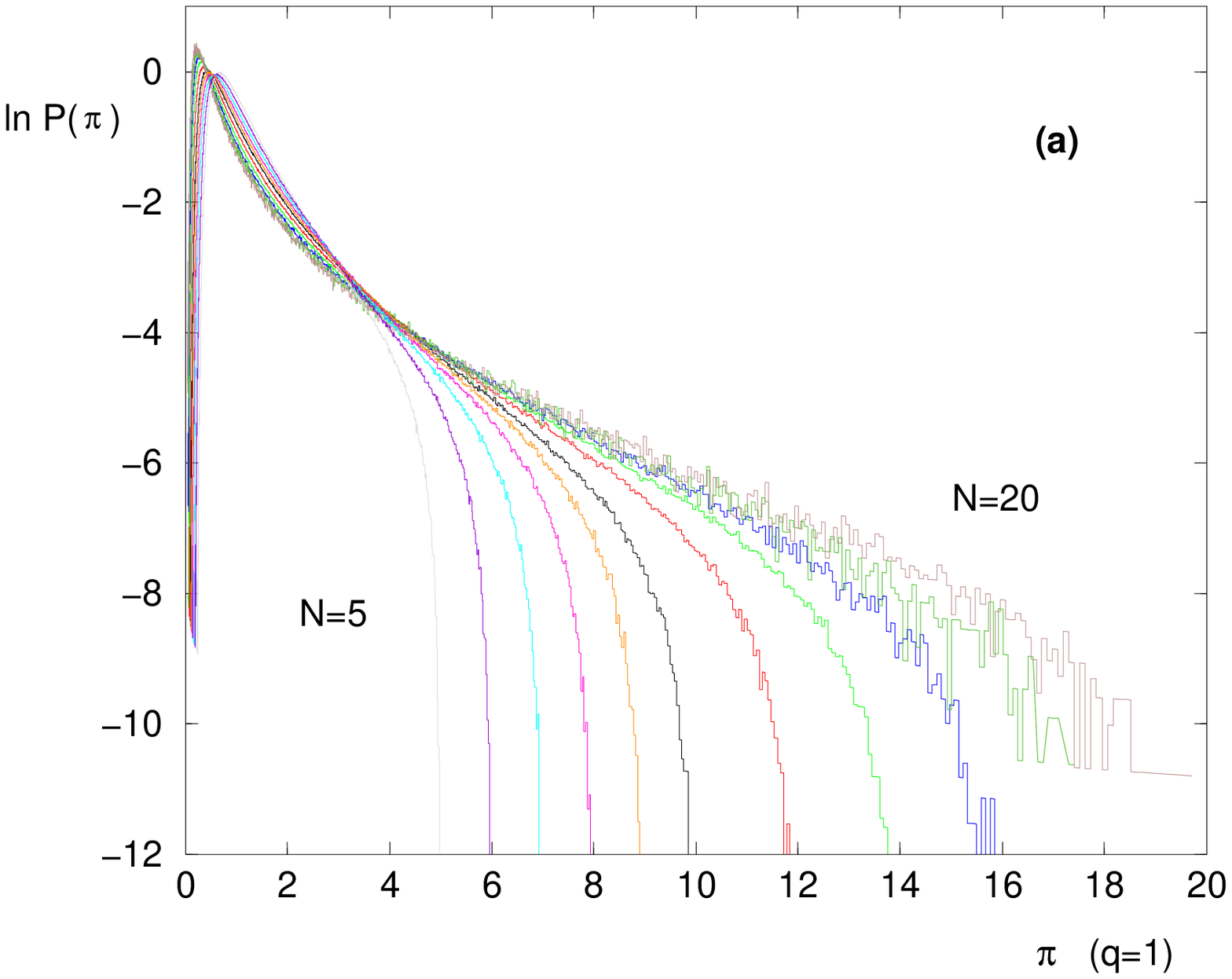}
\hspace{1cm}
 \includegraphics[height=6cm]{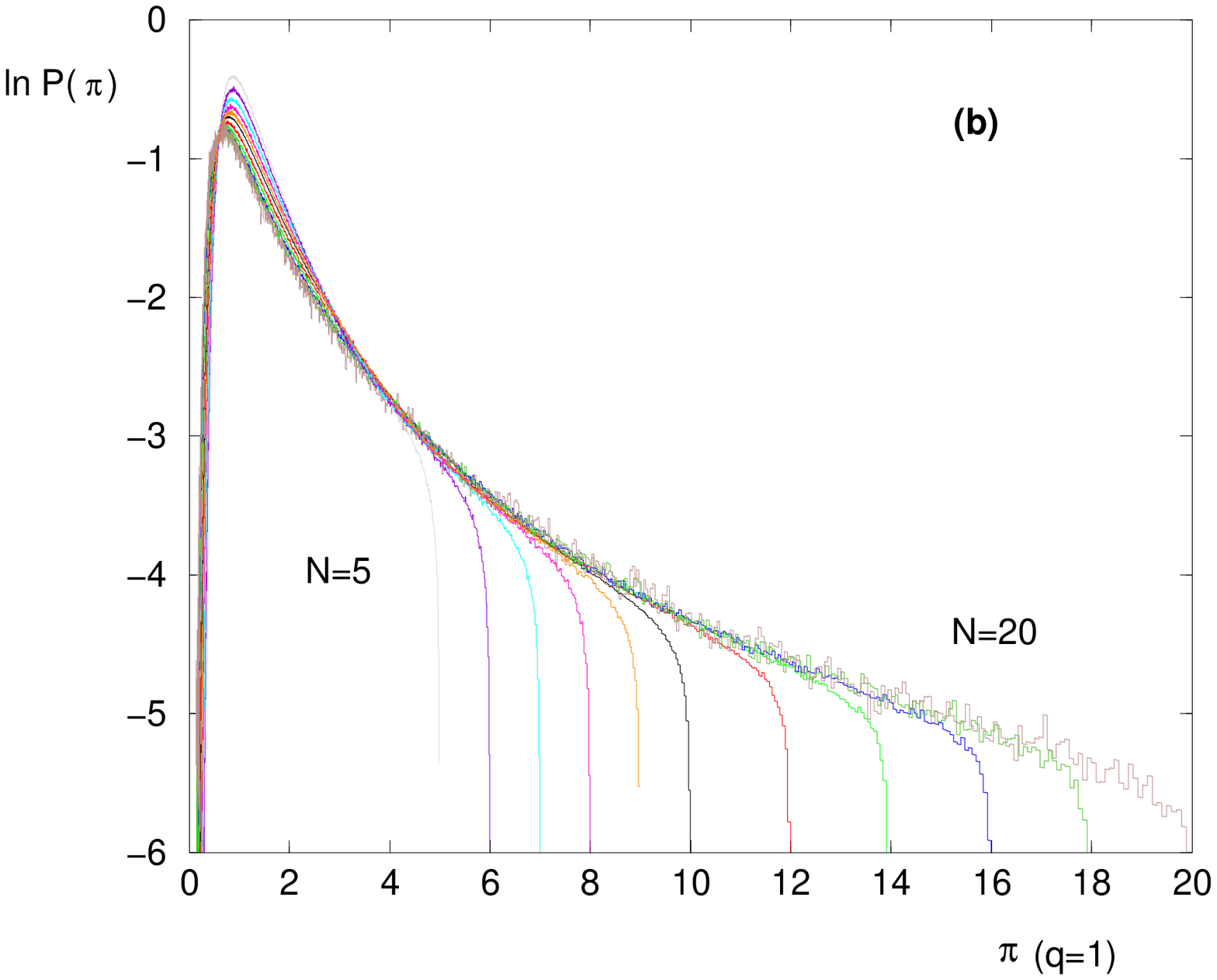}
\caption{ (Color on line) Statistics over the samples of the probability density $\pi(q=1)$
of full overlap between two configurations at criticality, for sizes  $N=5,6,7,8,9,10,12,14,16,18,20$ :
(a) Logarithm of the probability distribution $P( \pi(q=1) )$ for the DPCT
(b) Logarithm of the probability distribution $P( \pi(q=1) )$ for the REM  }
\label{histopiq1}
\end{figure}

By definition, the probability $P_N(N)$ of a full overlap $t=N$
coincides with the probability $Y_2=\sum w_i^2$ that the two walks
end at the same point
\begin{eqnarray}
 P_{T_c,N}(t=N)  \equiv Y_2 (N,T_c)
\label{pnndef}
\end{eqnarray}
 Using Eq. \ref{y2avcayleytc}, 
the average over the samples yields  
\begin{eqnarray}
{\overline P}_{T_c,N}(N)  \equiv \overline{ Y_2 }(N,T_c)
 \oppropto_{N \to \infty} \frac{1}{N} 
\label{pnn}
\end{eqnarray}
For the  disorder averaged overlap distribution of Eq. \ref{defpiq}, we thus obtain that
${\overline \pi}_{T_c,N}(q=1)$ remains finite as $N \to \infty$ :
\begin{eqnarray}
{\overline \pi}_{T_c,N}(q=1) \opsimeq_{N \to \infty} \overline{\pi}_{N=\infty}(q=1) >0
\label{piq1tc}
\end{eqnarray}
Beside the delta function $\delta(q)$ which bears the whole normalization
(Eq. \ref{piqtcav}), the asymptotic probability distribution
$\overline{\pi}_{N=\infty}(q)$ of the overlap $q$,
thus contains an isolated point at $q=1$ where
$\overline{\pi}_{N=\infty}(q=1) >0$.
This finite value at $q=1$ is due to rare events, since the typical value at $q=1$
is of order (Eq. \ref{yktypcayleytc})
\begin{eqnarray}
 \pi_{T_c,N}^{typ}(q=1) = N Y_2^{typ}(N,T_c) \oppropto_{N \to \infty} \frac{1}{N}
\label{piq1tctyp}
\end{eqnarray}
We show on Fig. \ref{histopiq1} a 
the probability $P_{T_c,N}( \pi(q=1) )$ over the samples of the probability
density $\pi_N(q=1)$ of full overlap between two configurations.
 From the probability of rare events with $Y_2 \to 1$
of Eq. \ref{ptcny2}, one obtains via the change of variables $\pi_N(q=1)=NY_2 $
the following singularity near the maximal value $\pi(q=1)  \to N $
\begin{eqnarray}
P_{T_c,N}(\pi(q=1) )  \opsimeq_{\pi(q=1)  \to N} \frac{ A_N}{N^{1+\sigma}}
 \left[ N -\pi(q=1)  \right]^{\sigma}
\label{ptcnpiq1tail}
\end{eqnarray}
where $A_N^{(DPCT)} \propto 1/N$ and $\sigma_{DPCT} \sim 1.5$.

\subsection{ Overlap distribution at criticality in the REM }

In the REM with $N$ spins, the spin overlap
\begin{eqnarray}
t=\sum_{i=1}^N S_i^{(1)}  S_i^{(2)}
\label{overspin}
\end{eqnarray}
can be defined from its relation with the $p$-spin glass model \cite{Der,Gro_Mez}
in the limit $p \to \infty$.
It is $t=N$ if the two configurations are identical ${\cal C}^{(1)}={\cal C}^{(2)}$
and $t < N$ if the two configurations are different  ${\cal C}^{(1)} \neq {\cal C}^{(2)}$.

As in Eq. \ref{pnn}, the probability $P_N(N)$ of a full overlap $t=N$
coincides with the probability $Y_2=\sum w_i^2$ that the two configurations
are the same. Using Eq. \ref{y2avremtc}, 
the average over the samples yields  
\begin{eqnarray}
{\overline P}_{T_c,N}^{REM}(N)  \equiv \overline{ Y_2 }(N,T_c)
 \oppropto_{N \to \infty} \frac{1}{N^{1/2}} 
\label{pnnrem}
\end{eqnarray}
For the overlap distribution of Eq. \ref{defpiq}, we thus obtain that
${\overline \pi}_{T_c,N}(q=1)$ diverges as $N \to \infty$ :
\begin{eqnarray}
{\overline \pi}_{T_c,N}^{REM}(q=1) \opsimeq_{N \to \infty} N^{1/2}
\label{piq1remtc}
\end{eqnarray}
Beside the delta function $\delta(q)$ which bears the whole normalization
(Eq. \ref{piqtcav}), the asymptotic probability distribution
$\overline{\pi}_{N=\infty}(q)$ of the overlap $q$,
thus contains an isolated point at infinity 
as a memory of the delta peak $(1-T/T_c) \delta(q-1)$ of the
low-temperature phase $T<T_c$.
Again this divergence at $q=1$ is due to rare events.
However here, in contrast with the directed polymer (Eq. \ref{piq1tctyp}),
the typical value at $q=1$
remains finite (Eq. \ref{yktypremtc})
\begin{eqnarray}
 \pi_{T_c,N}^{typ}(q=1) = N Y_2^{typ}(N,T_c) \simeq cst 
\label{piq1tctyprem}
\end{eqnarray}
We show on Fig. \ref{histopiq1} b 
the probability $P_{T_c,N}( \pi(q=1) )$ over the samples of the probability
density $\pi_N(q=1)$ of full overlap between two configurations.
The singularity near the maximal value is given by Eq. \ref{ptcnpiq1tail}
where $A_N^{(REM)} \propto 1/N^{1/2}$ and $\sigma_{REM} =0$.

\section{ Conclusion }

\label{conclusion}

In this paper, we have studied the weight statistics
at criticality for the Random Energy Model (REM)
and for the Directed Polymer on a Cayley Tree (DPCT)
with random bond energies.
These two mean-field disordered models present a freezing
transition with similar thermodynamic properties.
In particular, between the high temperature phase of extensive entropy
and the low-temperature phase of finite entropy, the entropy 
at criticality scales as $\overline{S}_N(T_c) \sim N^{1/2}$ in both models.
However, the statistical properties of the weights   
which coincide in the low-temperature phase become different
at the critical point. 
In the REM, all critical properties are governed 
by the finite-size exponent $\nu=2$ :
the typical values $e^{\overline{ \ln Y_k}}$ decay as $N^{-k/2}$, and 
the disorder-averaged values $\overline{Y_k}$
are governed by rare events and decay as $N^{-1/2}$ for any $k>1$.
In the DPCT, we find that the weight statistics is not governed
by the exponent $\nu=2$ of the thermodynamics,
but by another exponent $\nu'=1$ that had been
 previously mentioned in \cite{Coo_Der} 
in connection with finite-size corrections to the free-energy
below and at $T_c$. In particular, 	
the typical values $e^{\overline{ \ln Y_k}}$ decay  as $N^{-k}$,
and the disorder-averaged values $\overline{Y_k}$ decay as $N^{-1}$ for
any $k>1$. We have also presented numerical histograms
at criticality for the entropy, the maximal weight $w_{max}$
and $Y_2$. We have emphasized the role
of the rare samples that are still 'frozen' at $T_c$
( i.e. the rare samples having $S \sim 0$, $w_{max} \sim 1$, $Y_2 \sim 1$)
since it is the amplitude of these rare events that
governs the disorder averaged values $\overline{Y_k}$
as well as the overlap probability density ${\overline \pi}_{T_c,N}(q=1)$
of full overlap $q=1$.
 In particular, we have obtained that 
beside the delta function $\delta(q)$ which bears the whole normalization,  
 the disorder averaged asymptotic probability distribution ${\overline \pi}_{T_c,N=\infty}(q)$
contains an isolated point at $q=1$ as a memory of the delta peak $(1-T/T_c) \delta(q-1)$ of the
low-temperature phase $T<T_c$. The associated value
$\overline{\pi}_{N=\infty}(q=1)$ is finite for the DPCT, and diverges as
$\overline{\pi}_{N=\infty}(q=1) \sim N^{1/2}$ for the REM.

Concerning the weight statistics 
at criticality for the directed polymer,
let us finish by some comparison between the mean-field
version on the Cayley tree considered here
and the finite dimensional version
 that we have studied recently in \cite{DP3dmultif}.
We should first recall that in finite dimension $d$, the weights
of the $O(N^d)$ possible spatial positions of the polymer end-point
do not coincide with the configuration weights,
in contrast with the Cayley tree where the end-points are
in one-to-one correspondence with the $2^N$ configurations.
In finite dimension, the probability distributions
of the maximal weight $w_{max}$ and of $Y_2$
reach the values $w_{max}=1$ and $Y_2=1$
only for $T \leq T_{gap}$, where $T_{gap} < T_c$ \cite{DPweightslow},
whereas on the Cayley tree these two temperatures coincide $T_{gap}=T_c$.
This is why on the Cayley tree,
the disorder averaged values $\overline{ Y_k}$ for $k>1$ all decay
with a $k$-independent exponent $\overline{ Y_k} \propto 1/N$
representing the amplitude of rare events where $w_{max} \sim 1$,
whereas in finite dimension, the disorder averaged values $\overline{ Y_k}$ 
decay as $\overline{ Y_k} \propto 1/N^{ (k-1) \tilde D(k)}$ 
where the exponents $\tilde D(k)$ have a finite limit $\tilde D(+\infty)>0$.
Also in finite dimension, the comparison with
the exponents $D(k)$ governing the decay of typical values 
$Y_k^{typ}=e^{\overline{\ln  Y_k}} \propto 1/N^{ (k-1)  D(k)}$
show that the threshold $k_c$ between the region
$k \leq k_c$ where they coincide $D(k)=\tilde D(k)$ 
and the region $k > k_c$ where they differ $D(k) > \tilde D(k)$
is of order $k_c \sim 2$ \cite{DP3dmultif}, whereas
on the Cayley tree, the exponents for averaged and typical values
are always different as soon as $k>1$.
So the role of rare events is stronger 
on the Cayley tree.

\appendix

\section{L\'evy sums for $\mu<1$}

\label{app_mu}

In this Appendix, we recall some properties of L\'evy sums with $\mu<1$,
since their weight statistics is the same as in the Random Energy Model
in the low-temperature phase with $\mu(T)=T/T_c$.

\subsection{ Weight statistics in L\'evy sums}

The sum 
\begin{equation}
\Sigma_M = \sum_{i=1}^M x_i 
\label{sum}
\end{equation}
of $M$ positive independent variables $(x_1,..x_M)$ distributed with
a probability distribution that decays algebraically
\begin{equation}
\rho(x) \opsimeq_{x \to +\infty} \frac{A}{x^{1+\mu} }
\label{levymu}
\end{equation}
has very special property when $0<\mu<1$ since the first moment diverges
$<x>=+\infty$ \cite{Der,levy} : the sum $\Sigma_M$ grows as $M^{1/\mu}$,
and the rescaled variable is distributed with a stable L\'evy distribution.
Another important property is that the maximal variable $x_{max}(M)$
among the $M$ variables $(x_1,...x_M)$ is also of order $M^{1/\mu}$,
i.e. the sum $\Sigma_M$ is actually dominated by the few biggest terms.
To quantify this effect, it is convenient to introduce the weights
\begin{equation}
w_i = \frac{x_i}{\Sigma_M}
\end{equation}
and their moments
\begin{equation}
Y_k=\sum_{i=1}^M w_i^k
\end{equation}

The link with the weight statistics in the Random Energy Model
can be understood as follows.
The lowest energy in the REM
is distributed exponentially 
\begin{eqnarray}
P_{extremal}(E) \opsimeq_{E \to -\infty} e^{ \gamma E}
\label{tailexp}
\end{eqnarray}
This exponential form that corresponds to the tail of the Gumbel distribution
for extreme-value statistics \cite{Gum_Gal,Bou_Mez},
immediately yields that the Boltzmann weight $x=e^{-\beta E}$ 
has a distribution that decays algebraically (Eq. \ref{levymu})
with exponent
\begin{eqnarray}
\mu= T \gamma
\end{eqnarray}
In the REM, the coefficient $\gamma$ in the exponential (Eq. \ref{tailexp})
is $\gamma=1/T_c$.

Let us also mention that in the mean-field SK model
of spin-glasses, exactly the same expressions of $Y_k$ (Eq. \ref{yklevy})
also appear \cite{Me_Pa_Vi,replica}, but with a different 
interpretation : the weights are those of the pure states.
As a consequence, the parameter $\mu(T)$ which is a complicated function
of the temperature vanishes at the transition $\mu(T_c)=0$
(only one pure state in the high temperature state)
and grows at $T$ is lowered towards 
$\mu(T=0)$ of order $ 0.5$ \cite{SKzerotemp}.
This is in contrast with the REM model where $\mu(T)=T/T_c$
grows with the temperature from $\mu(T=0)=0$ (only one ground state)
to $\mu(T_c)=1$ at the transition, where the number 
of important microscopic states is not finite anymore.
Nevertheless, the expression (Eq. \ref{yklevy}) for the weights
of pure states means that the free-energy $f$ of pure states in the
SK model is distributed exponentially
\begin{eqnarray}
P(f) \opsimeq_{f \to -\infty} e^{ \gamma(T) f}
\end{eqnarray}
with a parameter $\gamma(T)= \mu(T)/T$.

\subsection{ Disorder-averaged moments $\overline{Y_k}^{Levy} $ }

The averaged values in the limit $M \to \infty$
are finite for $0<\mu<1$ and reads \cite{Der} 
\begin{equation}
\overline{Y_k}^{Levy}= \frac{\Gamma(k-\mu)}{\Gamma(k) \Gamma(1-\mu)}
\label{yklevy}
\end{equation}
Let us recall how one derives this result \cite{Der},
since it will be useful for the critical case $\mu_c=1$
considered in Appendix \ref{app_muc}.
It is convenient to exponentiate the denominator according to \cite{Der}
\begin{equation}
Y_k = \frac{1}{ \Gamma(k)} \int_0^{+\infty} dt t^{k-1} e^{-t \sum_{i} x_i^k} \sum_j x_j^k 
\label{yklevy1}
\end{equation}
in order to perform the average
\begin{equation}
\overline{Y_k}^{Levy}= \frac{M}{ \Gamma(k)} \int_0^{+\infty} dt t^{k-1} 
\ \ \overline{ x^k e^{-t x} } \ \ ( \overline{e^{-t x} } )^{M-1} 
\label{yklevy2}
\end{equation}
For large $M$, the integral will be dominated by the region where $t$
is small, and one may approximate \cite{Der} 
\begin{equation}
\overline{ x^k e^{-t x} } = \int dx \rho(x) x^k e^{-t x} \sim A t^{\mu -k} \Gamma(k-\mu)
\label{yklevy3}
\end{equation}
and
\begin{equation}
 \overline{e^{-t x} } = \int dx \rho(x) e^{-t x} \sim e^{ -t^{ \mu} A (- \Gamma(-\mu))}
\label{yklevy4}
\end{equation}
yielding
\begin{equation}
\overline{Y_k}^{Levy}= \frac{M A \Gamma(k-\mu)}{ \Gamma(k)} \int_0^{+\infty} dt t^{\mu-1} 
e^{ - M t^{ \mu} A (- \Gamma(-\mu))}
\label{yklevy5}
\end{equation}
leading to Eq. \ref{yklevy} in the limit $M \to \infty$.
More generally, correlations functions between $Y_k$ can also
be computed \cite{Der},  in particular
\begin{equation}
\overline{Y_k^2}^{Levy}= \frac{1}{\Gamma(2 k)} \left( 
\frac{\Gamma(2 k-\mu)}{ \Gamma(1-\mu)}
+ \mu \frac{\Gamma^2 (k-\mu)}{ \Gamma^2(1-\mu)}
\right) 
\label{yk2levy}
\end{equation}

\subsection{ `Typical values'
 $ Y_k^{Levy}(typ) = e^{ \overline{\ln Y_k} }$ }

To compute the disorder average of the logarithm of $Y_k$,
we first rewrite
\begin{equation}
 \ln Y_k= \ln \left( \sum_{i=1}^M x_i^k \right) - k \ln \left( \sum_{i=1}^M x_i \right)
\label{lnyklevy}
\end{equation}
and use the identity
\begin{equation}
\ln Z = \int_0^{+\infty} \frac{dt}{t} \left( e^{-t} - e^{ - t Z} \right)
= \lim_{\epsilon \to 0} \left[  \int_0^{+\infty} dt \ t^{ \epsilon-1} e^{-t} 
-  \int_0^{+\infty} dt \ t^{ \epsilon-1} e^{-t Z } \right]
\label{identity}
\end{equation}
Using Eq \ref{yklevy4}, we obtain 
\begin{equation}
\overline{ \ln \left( \sum_{i=1}^M x_i \right) }  = \int_0^{+\infty}  
 \frac{dt}{t} \left[ e^{-t} -  \left(  \overline{e^{-t x} }  \right)^M \right]
 \sim \left( 1-\frac{1}{\mu} \right) \Gamma'(1) +
\frac{1}{\mu} \ln \left( M A (-\Gamma(-\mu)) \right) 
\label{lnyklevy1}
\end{equation}
and similarly
\begin{equation}
\overline{  \ln \left( \sum_{i=1}^M x_i^k \right) } = \int_0^{+\infty}  
 \frac{dt}{t} \left[ e^{-t} -  \left(  \overline{e^{-t x^k} }  \right)^M \right]
 \sim  \left( 1-\frac{k}{\mu} \right) \Gamma'(1) +
\frac{k}{\mu} \ln \left( M \frac{A}{k}  (-\Gamma(- \frac{\mu}{k} )) \right) 
 \label{lnyklevy2}
\end{equation}
so that finally in the limit $M \to \infty$ (Eq. \ref{lnyklevy})
\begin{equation}
\overline{ \ln Y_k}= (1-k) \Gamma'(1) 
+ \frac{k}{\mu} \ln \left( \frac{\Gamma(1 - \frac{\mu}{k} )}{\Gamma(1-\mu)} \right) 
\label{lnyklevyres}
\end{equation}

\subsection{ Critical behaviors near the transition point $\mu \to \mu_c=1$ }

As $\mu \to 1$,  Eq. \ref{lnyklevyres} gives the following leading term
\begin{equation}
\overline{ \ln Y_k} \sim k \ln (1-\mu) 
\label{lnyklevymuto1}
\end{equation}
i.e. the typical values vanish as
\begin{equation}
Y_k^{Levy}(typ) = e^{ \overline{\ln Y_k} }   \sim (1-\mu)^k
\label{typyklevymuto1}
\end{equation}
whereas the averaged moments of Eq. \ref{yklevy} vanish linearly as
\begin{equation}
\overline{Y_k}^{Levy} = \frac{(1- \mu)}{ k-1 }  +O((1-\mu)^2)
\label{yklevymuto1}
\end{equation}
as well as higher moments, for instance Eq. \ref{yk2levy}
\begin{equation}
\overline{Y_k^2}^{Levy} \sim  \frac{(1- \mu)}{ 2 k-1 }+O((1-\mu)^2)
\label{yk2levymuto1}
\end{equation}
This shows that disorder-averaged values are governed by the rare events
where the maximal weight $w_{max}$ is near $1$ : 
the density $f(w)$ of Eq. 
\ref{densitew} becomes for $\mu \to 1$
\begin{equation}
f_{Levy}(w) \opsimeq_{\mu \to 1} (1-\mu) w^{-1-\mu} (1-w)^{\mu-1} 
\label{densitewmu}
\end{equation}

\section{ Weight statistics for L\'evy sums at criticality $\mu_c=1$  }

\label{app_muc}

In this Appendix, we describe some results on 
the weight statistics for L\'evy sums at criticality $\mu_c=1$
to compare with the results given in the text 
for the Random Energy Model and
for the Directed Polymer on a Cayley Tree.
For $\mu_c=1$, the sum $\Sigma_M$ of Eq. \ref{sum} scales as $M \ln M$,
whereas the maximal value $x_{max}$ among the $M$ variables
scales as $M$ \cite{levy} :  the decay of the $Y_k$ is thus expected to 
depend on the variable $(\ln M)$.

\subsection{ Decay of disorder averaged values $\overline{Y_k}(M) $ }

We start from Eq. \ref{yklevy2}
\begin{equation}
\overline{Y_k}= \frac{M}{ \Gamma(k)} \int_0^{+\infty} dt t^{k-1} 
\ \ \overline{ x^k e^{-t x} } \ \ ( \overline{e^{-t x} } )^{M-1} 
\label{yklevy2bis}
\end{equation}
For large $M$, the integral will be dominated by the region where $t$
is small, and one may approximate
\begin{equation}
\overline{ x^k e^{-t x} } = \int dx \rho(x) x^k e^{-t x} 
\sim A t^{1 -k} \Gamma(k-1)
\label{yklevy3bis}
\end{equation}
and
\begin{equation}
 \overline{e^{-t x} } = \int dx \rho(x) e^{-t x}
 \sim e^{ - A t \ln \frac{1}{t}}
\label{yklevy4bis}
\end{equation}
yielding 
\begin{equation}
\overline{Y_k}=
\frac{M A }{ k-1 } \int_0^{+\infty} dt  
 \  e^{ - M  A t \ln \frac{1}{t}} \sim \frac{1}{ (k-1)  \ln M} 
\label{ykavcriti}
\end{equation}

\subsection{ Disorder averaged entropy   }

From Eq. \ref{yklevy2bis},
the disorder-averaged entropy (Eq. \ref{entropyGM} )reads
\begin{eqnarray}
\overline{S(M)} 
= -\partial_k \ \overline{Y_k}  \vert_{k=1} 
= M  \int_0^{+\infty} dt  
  \ ( \overline{e^{-t x} } )^{M-1}  \   \int dx \rho(x) 
\left[ (\Gamma'(1) -\ln t -\ln x)  x e^{-t x} \right] 
\label{slevy1}
\end{eqnarray}
Using Eq. \ref{yklevy4bis}, one obtains at leading order
\begin{equation}
\overline{S(M)} \propto \ln ( M A \ln M) 
\label{slevyres}
\end{equation}

\subsection{ Decay of typical values
 $ Y_k^{typ}(M) = e^{ \overline{\ln Y_k} }$  }

To compute the disorder average of the logarithm of $Y_k$,
we start from Eqs \ref{lnyklevy} and \ref{identity}.
Using Eq \ref{yklevy4bis}, we obtain 
\begin{equation}
\overline{ \ln \left( \sum_{i=1}^M x_i \right) } = \int_0^{+\infty}  
 \frac{dt}{t} \left[ e^{-t} -  \left(  \overline{e^{-t x} }  \right)^M \right]
\sim  \int_0^{+\infty}  
 \frac{dt}{t} \left( e^{-t} - e^{ - M A t \ln \frac{1}{t}} \right)
 \sim \ln ( M A \ln M )
\label{lnyklevy1bis}
\end{equation}
and as in Eq \ref{lnyklevy2} with $\mu=1$
\begin{equation}
\overline{  \ln \left( \sum_{i=1}^M x_i^k \right) } = \int_0^{+\infty}  
 \frac{dt}{t} \left[ e^{-t} -  \left(  \overline{e^{-t x^k} }  \right)^M \right]
 \sim  \left( 1- k \right) \Gamma'(1) +
k \ln \left( M \frac{A}{k}  (-\Gamma(- \frac{1}{k} )) \right) 
 \label{lnyklevy2bis}
\end{equation}
so that finally  the leading term is (Eq. \ref{lnyklevy})
\begin{equation}
\overline{ \ln Y_k}=
\overline{ \ln \left( \sum_{i=1}^M x_i^k \right) }
 - k \overline{ \ln \left( \sum_{i=1}^M x_i \right) } \sim -k \ln ( \ln M)
\label{lnyklevyresbis}
\end{equation}
i.e 
\begin{equation}
Y_k^{typ}(M) = e^{ \overline{\ln Y_k} } \sim \frac{1}{ (\ln M)^k }
\label{lnyklevyrestyp}
\end{equation}

\subsection{ Finite-size scaling in the critical region }

The comparison of the results for the entropy, 
for the disorder-averaged values 
and for the typical values of the $Y_k$
between the phase $\mu<1$ and the critical point $\mu_c=1$
shows that the appropriate scaling variable is $(1-\mu) \ln M $, corresponding to
a 'finite-size exponent'
\begin{equation}
\nu_{Levy}=1
\end{equation}
This is in contrast with the Random Energy Model where the number of configurations is $M=2^N$,
and the appropriate finite-size scaling behavior (Eq. \ref{nu}) is
 $(T_c-T) N^{1/2} = (1-\mu) ( \ln M)^{1/2}$ with $\nu_{REM}=2$.

\subsection{ Probability distributions of $w_{max} $ and of $Y_2$  }

\begin{figure}[htbp]
\includegraphics[height=6cm]{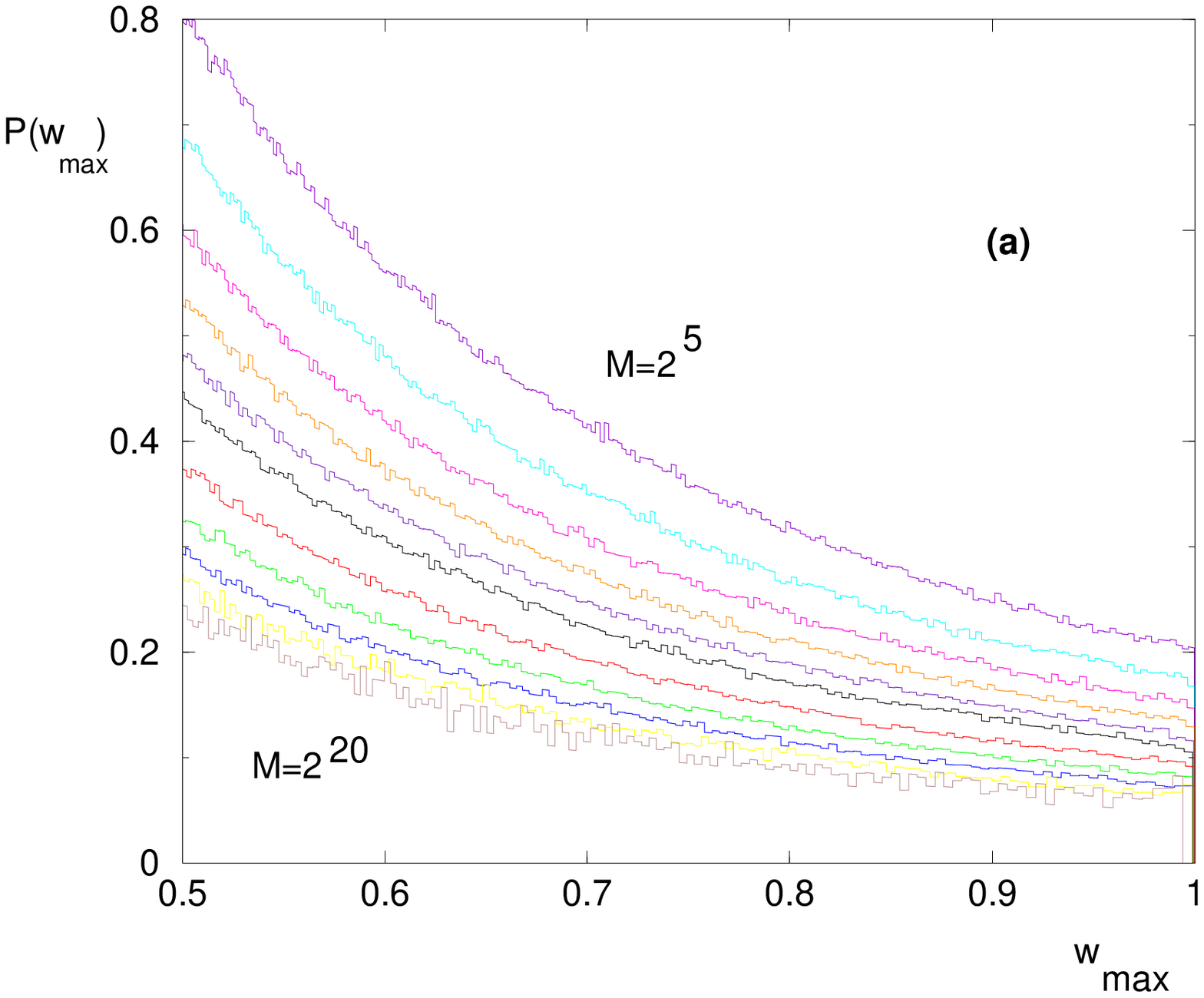}
\hspace{1cm}
\includegraphics[height=6cm]{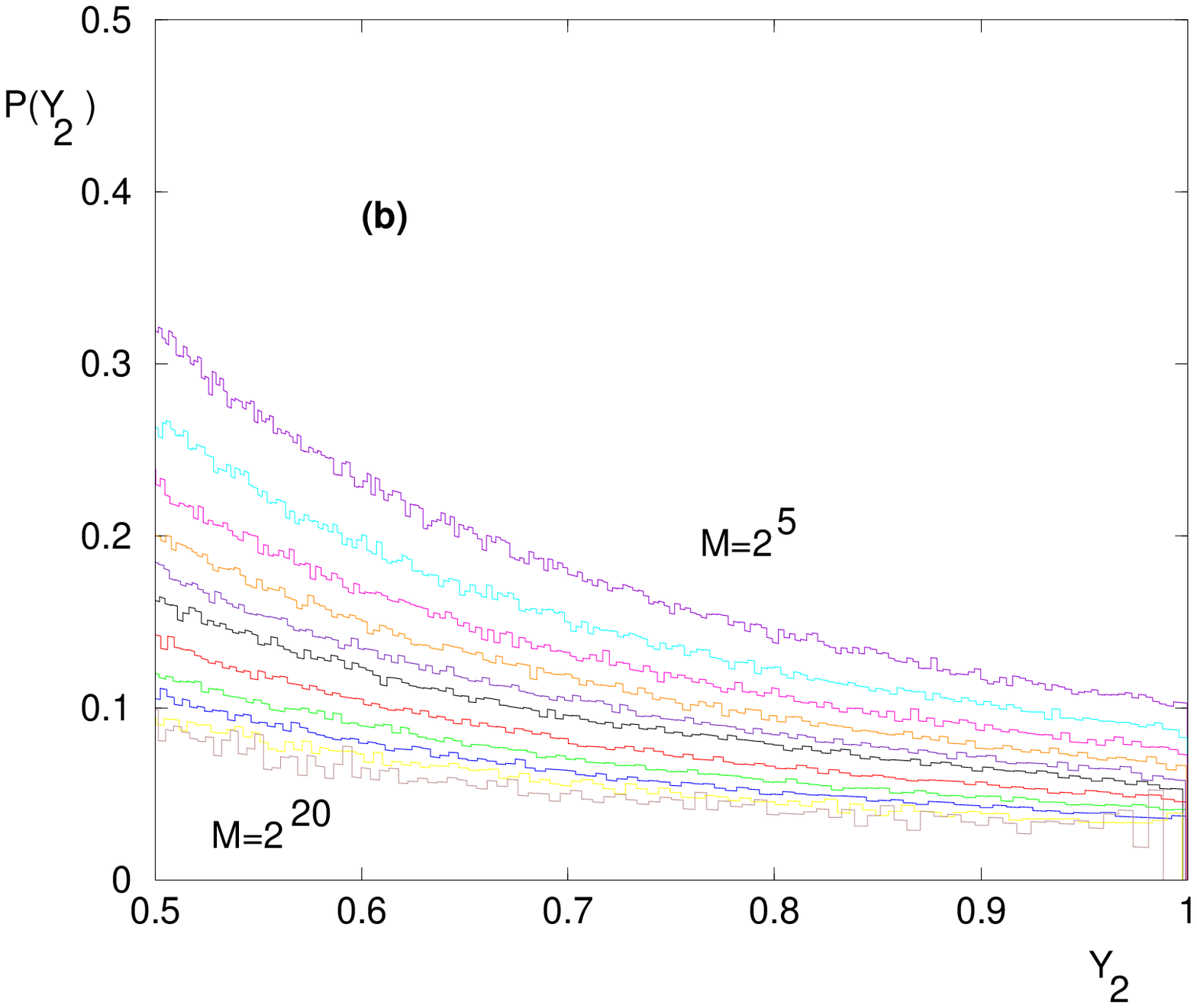}
\caption{ (Color on line) Weight statistics in L\'evy sums of $M$ terms
( with $2^5 \leq M \leq 2^{20}$ ) for the critical value $\mu_c=1$ : 
(a) Probability distribution of the maximal weight
 $w_{max}$ in the region $1/2 \leq w_{max} \leq 1$
 (b) Probability distribution of $Y_2$ 
in the region $1/2 \leq Y_2 \leq 1$   }
\label{rarelevyhisto}
\end{figure}

We show on Fig. \ref{rarelevyhisto} the finite-size probability distributions
of the maximal weight $w_{max}$ and $Y_2$ at the critical value
$\mu_c=1$, to compare with the corresponding figures
given in the text for the REM (Fig. \ref{rareremhisto})
and for the DPCT (Fig. \ref{rarecayleyhisto})
with the correspondence $M=2^N$.
As in Eq. \ref{ptcnwmax}, the behavior of
 the probability distribution $P_{\mu_c,M}(w_{max})$ near $w_{max} \to 1$
for finite sums of $M$ terms at the critical value $\mu_c=1$
is of the form
\begin{eqnarray}
P_{\mu_c,M}(w_{max}) \opsimeq_{w_{max} \to 1} A_M (1-w_{max})^{\sigma}
\end{eqnarray}
The amplitude $A_M$ of these rare events
is the amplitude that governs the disorder averaged values $\overline{Y_k}$
of Eq. \ref{ykavcriti}
\begin{eqnarray}
A_M \oppropto_{M \to \infty} \frac{1}{ \ln M}
\end{eqnarray}
The singularity exponent $\sigma$ is simply
\begin{eqnarray}
\sigma_{Levy}=0
\end{eqnarray}
 in continuity with the rare events in the region $\mu<1$.
This value is the same as in the REM  (Eq. \ref{sigmarem})
but different from the value measured in the DPCT (Eq. \ref{sigmacayley}).

\subsection{ Conclusion : comparison with the REM and the DPCT}

In this Appendix, we have described the weight statistics in L\'evy sums
for the critical value $\mu_c=1$, to compare with the REM and the DPCT
cases studied in the text. Although the three models have the same properties 
in the low-temperature phase with $\mu=T/T_c<1$, 
we find here that the three models have different critical finite-size properties.
The REM and the L\'evy sums involve a single finite-size exponent
\begin{eqnarray}
\nu_{REM} && =2 \\
\nu_{Levy} && =1 
\end{eqnarray}
and both have a singularity exponent
\begin{eqnarray}
\sigma_{REM} = 0  = \sigma_{Levy}  
\end{eqnarray}
 in continuity with its low-temperature value $\sigma=\mu-1 \to 0$.
On the contrary, the DPCT involves two exponents 
\begin{eqnarray}
\nu_{DPCT} && =2 \\
\nu_{DPCT}' && =1 
\end{eqnarray}
The exponent $\nu=2$ governs the thermodynamics, in particular the entropy
and the specific heat, whereas $\nu'=1$ governs the $Y_k$ statistics.
Moreover, the singularity exponent at criticality
\begin{eqnarray}
\sigma_{DPCT} \simeq 1.5  
\end{eqnarray}
is very different from the limit of its low-temperature value $\sigma=T/T_c-1 \to 0$.

\end{document}